\documentclass[10pt,conference]{IEEEtran}
\IEEEoverridecommandlockouts
\usepackage{cite}
\usepackage{amsmath,amssymb,amsfonts}
\usepackage{graphicx}
\usepackage{textcomp}
\usepackage{xcolor}
\usepackage{booktabs}
\usepackage{multirow}
\usepackage{float}
\usepackage{quantikz}
\providecommand{\setwiretype}[1]{}
\usepackage{subcaption}
\usepackage[hyphens]{url}
\usepackage[hidelinks]{hyperref}

\graphicspath{{figures/}}

\def\BibTeX{{\rm B\kern-.05em{\sc i\kern-.025em b}\kern-.08em
    T\kern-.1667em\lower.7ex\hbox{E}\kern-.125emX}}

\begin{document}
\raggedbottom

\title{Quantum PDE Solvers in Practice: Application-Driven Benchmarking of the Heat Equation}

\author{%
\IEEEauthorblockN{Mahmoud ElKarargy\textsuperscript{1},
Abdelaziz Rahwan\textsuperscript{1},
Abdelrahman Elsayed\textsuperscript{1},
Forat Hatem\textsuperscript{1}}
\IEEEauthorblockA{\textsuperscript{1}Brightskies Technologies, Alexandria, Egypt\\
\{mahmoud.elkarargy, abdelaziz.rahwan\}@brightskiesinc.com\\
\{abdo.elsayd102, forathatem2004\}@gmail.com}
}

\maketitle

\begin{abstract}
Quantum PDE solvers are difficult to evaluate in practice because the
literature compares algorithms under different discretisations, output
models, reconstruction rules, and hardware assumptions.

This paper contributes a reproducible, application-driven benchmark for
the 1-D Dirichlet heat equation that places eleven kernels from five
paradigm classes under the same problem instance and readout contract:
\textbf{(A)} coherent linear solvers (HHL, QSVT, QLS-Fourier),
\textbf{(B)} a NISQ variational linear solver (VQLS),
\textbf{(C)} imaginary-time variational dynamics (QITE, var-QITE,
AVQDS), \textbf{(D)} real-time Hamiltonian simulation and unitary
dilations (Hamiltonian simulation, Schade-Hamiltonian,
Schr\"odingerisation), and \textbf{(E)} a spectral/transform method
(QSM).

The harness fixes three initial conditions (pulse, Gaussian, bimodal),
four grid scales ($n{=}4$--$7$ qubits, $N{=}16$--$128$), a CFL-like
ratio $r{\approx}0.4$, and a final time $T{=}1$. Three backends
(statevector, ideal shots at $10^{5}$/step, and noisy Aer) isolate
algorithmic, sampling, and device-noise error; backend-specific kernel
omissions are stated in \S\ref{sec:backends}.

The central finding is that QSM and Schade-Hamiltonian reproduce the
semi-discrete reference to floating-point precision on statevector;
Schr\"odingerisation reaches $\sim10^{-4}$; and QITE is the strongest
non-transform method on smooth data. Conversely, in our fixed-shot
setting, HHL degrades to $\sim0.79$ relative $\ell_2$ error, and several
low-depth or postselected methods become readout-limited before their
unitary cores are exhausted. A norm-mismatch ablation shows that the
shared smooth-initial-condition error plateau of Hamiltonian simulation,
AVQDS, and QLS-Fourier is partly a reconstruction-normalisation artifact
(23--29\% of their $n{=}7$ smooth-IC error). Finally, compact observables
such as total thermal energy and individual Fourier-mode weights require
1--3 orders of magnitude fewer shots than full-field reconstruction on
the same circuits.

The result is a public benchmark artifact and a practical selection
guide: spectral/dilation kernels for reference accuracy, QITE for
moderate-depth smooth full-field studies, and compact observable
extraction for credible application-level quantum benefit.
\end{abstract}

\begin{IEEEkeywords}
quantum computing, quantum algorithms, partial differential equations, 
heat equation, benchmarking, Hamiltonian simulation, Schr\"{o}dingerisation, quantum linear systems,
variational quantum algorithms, quantum singular value transformation,
NISQ
\end{IEEEkeywords}

\section{Introduction}
\label{sec:introduction}

Partial differential equations (PDEs) underpin models across physics,
engineering, and the computational
sciences~\cite{evans2010partial}, and within the quantum-algorithms
literature have emerged as an important application domain because
they exercise state preparation, evolution, inversion, and measurement
in a single problem. Yet that literature is difficult to compare:
methods differ in discretisation, computational mechanism, output
representation, and hardware assumption, making it unclear how to
relate conceptual promise to practical performance or how to interpret
quantum-advantage claims on a common basis.

Among candidate benchmark problems, the heat equation provides a
particularly useful lens through which to organize this landscape. In
its general parabolic form, it may be written as
\begin{equation}
  \frac{\partial u}{\partial t}
  = \alpha \,\Delta u,
  \qquad \mathbf{x}\in\Omega \subset \mathbb{R}^d,\; t>0,
  \label{eq:heat_general}
\end{equation}
where $\Delta$ denotes the Laplacian operator and $\alpha>0$ is the
diffusion coefficient. This form captures the broader class of
diffusion-type dynamics relevant to parabolic PDEs. In this paper, we
specialize to the one-dimensional case with Dirichlet boundary
conditions,
\begin{equation}
  \frac{\partial u}{\partial t}
  = \alpha\,\frac{\partial^2 u}{\partial x^2},
  \quad x\in(0,L),\; t>0,
  \label{eq:heat}
\end{equation}
which is analytically tractable and simple enough to support controlled
comparison across multiple quantum algorithms.  Three properties make
this PDE unusually well-suited as a cross-paradigm benchmark: (i)~it
exercises all four primitive operations that distinguish quantum PDE
solvers from one another---state preparation, matrix inversion,
real- and imaginary-time evolution, and amplitude readout, within one
instance; (ii)~its closed-form Fourier sine-series solution makes the frequency
content of each initial condition explicit, enabling initial condition (IC) based
stress-testing of kernels that differ in their handling of high-mode
dynamics (pulse excites a broad band; Gaussian is low-frequency
dominated; bimodal isolates two discrete modes); and (iii)~representative quantum treatments of the heat
equation~\cite{linden2022quantum,jin2024prl,jin2024schrodingerisation,alipanah2025avqds,tseng2025comparative},
together with broader linear parabolic PDE
solvers~\cite{kumar2024qitepde}, align this benchmark with active
lines of research and a ready-made comparison literature.  Together these properties make
the heat equation a natural common reference problem for examining how
different quantum paradigms approach PDEs.

Building on Feynman's observation~\cite{feynman1982simulating} that
quantum systems obey differential equations, the
Harrow--Hassidim--Lloyd algorithm~\cite{harrow2009quantum} opened a
pathway to quantum linear-system methods for implicit PDE
discretisations. Subsequent work has produced improved linear-system
and differential-equation methods, Hamiltonian-based formulations,
and general PDE simulation frameworks~\cite{berry2014high,childs2021high,costa2019quantum,linden2022quantum,jin2024prl,jin2023schrodingerization};
in parallel, variational approaches~\cite{cerezo2021variational,bravo2023variational}
offer hardware-accessible alternatives for the NISQ
era~\cite{preskill2018nisq}. What is missing from this landscape is an
empirically grounded comparison of these approaches across a common
problem setting.

\textbf{This paper} asks a concrete empirical question: \emph{Under
matched discretisation, solution reconstruction, and shot-budget
assumptions, which quantum kernels accurately solve the heat
equation, and what determines their characteristic failure modes?} We
implement eleven kernels spanning five paradigm classes in a single
harness, evaluate them on three initial conditions, and compare
statevector, ideal shot-based, and noisy Aer backends
(\S\ref{sec:backends}). To our knowledge, no prior benchmark spans all
five paradigm classes on a shared PDE instance under a fixed
reconstruction rule; the closest comparative studies each cover a
strict subset of kernels, backends, or output maps
(\S\ref{sec:related}). We extend this line by adding Paradigm~A
coherent solvers, Paradigm~E spectral methods, the
ideal-shots/noisy-simulator backend ladder, and a norm-mismatch
ablation that decomposes the shared smooth-IC error plateau into its
algorithmic and reconstruction components, all under one output map.

\textbf{Key contributions.}

\textbf{(C1)}~A controlled eleven-kernel benchmark spanning five
paradigms under one PDE instance, discretisation, and reconstruction
rule (\S\ref{sec:kernels}--\S\ref{sec:comparison}), with a staged
statevector$\!\to\!$ideal-shots$\!\to\!$noisy-Aer backend ladder that
separates algorithmic, sampling, and device-noise error
(\S\ref{sec:backends}).

\textbf{(C2)}~A norm-mismatch ablation that attributes the smooth-IC
error plateau shared by Hamiltonian simulation, AVQDS, and QLS-Fourier
to a reconstruction-normalisation failure, with 23--29\% of the
residual $\ell_{2}$ error at $n{=}7$ traced to pure norm drift
independent of the unitary core (\S\ref{sec:norm-ablation}). This
reframes several apparent algorithmic ceilings as readout-pipeline
artefacts and explains why kernels with very different mechanisms can
produce the same error band.

\textbf{(C3)}~An observable-readout advantage analysis
(\S\ref{sec:observable-advantage}) showing that compact physical
functionals (total energy, selected mode weights, and boundary-flux
proxies) reach target precision at $1$--$3$ orders of magnitude lower
shot budget than full-field reconstruction on the same circuits.

\textbf{(C4)}~A reproducibility-oriented benchmark artifact: a modular
framework, configuration files, raw outputs, and scripts for regenerating
the tables and figures, with new kernels added through one documented
interface method.

\section{Related Work}
\label{sec:related}

\textbf{Prior comparative studies} each cover a strict subset of the
landscape. Tseng et al.~\cite{tseng2025comparative} compare two
heat-equation solvers at statevector level;
Shayegan~\cite{shayegan2026qlscomp} benchmarks VQLS, HHL, and quantum
annealing on time-fractional diffusion;
\"Ozg\"uler~\cite{ozguler2025qpdebench} evaluates VQE against VarQTE-,
AVQDS-, and Trotter-based kernels in a shared harness on
advection--diffusion; Alipanah et al.~\cite{alipanah2025avqds} compare
AVQDS and var-QITE against a Trotter baseline;
surveys~\cite{morales2024qlsp,tennie2025nrpnonlinear,dalzell2023quantum}
cover quantum linear solvers and the broader PDE landscape without
implementing kernels.
None of these span all five paradigms, combine
statevector/shot/noisy backends, enforce a shared reconstruction
rule, or quantify the reconstruction component of the measured error.

\textbf{Kernel-specific literature.}
Paradigm~A builds on HHL~\cite{harrow2009quantum}, QSVT block
encodings~\cite{gilyen2019quantum,martyn2021grand,kharazi2025block}
(benchmarked against HHL in~\cite{lefterovici2026functional}), and the
Fourier-LCU construction~\cite{childs2017quantum}; our fixed-CFL
condition-number analysis (\S\ref{sec:cond-and-scaling}) shows $A$ is
well-conditioned at all tested $n$, so we do not engage the
amplitude-amplification~\cite{ambainis2012variable} and
preconditioning~\cite{clader2013preconditioned} refinements developed
for the ill-conditioned regime. Paradigm~B follows
Bravo-Prieto~\cite{bravo2023variational} with dynamic-ansatz and
finite-element extensions~\cite{patil2022dynvqls,trahan2023vqls} and
related Poisson VQAs~\cite{liu2021poisson}. Paradigm~C builds on
Motta~\cite{motta2020determining}, extended to linear PDEs by Kumar
and Wilmott~\cite{kumar2024qitepde}, with McLachlan variational
dynamics~\cite{mclachlan1964variational,yuan2019theory} and the AVQDS
construction of Alipanah~et~al.~\cite{alipanah2025avqds}.
Paradigm~D combines
Schr\"{o}dingerisation~\cite{jin2024prl,jin2024schrodingerisation,jin2023schrodingerization}
with the Hamiltonian IBVP construction of
Schade~et~al.~\cite{schade2023hamiltoniansim} and mean-field
Trotterisation from~\cite{alipanah2025avqds}. Paradigm~E follows the
DST primitives of Klappenecker and
R\"otteler~\cite{klappenecker2001discrete} as specialised for
non-periodic BVPs in Febrianto~et~al.~\cite{febrianto2025spectral}.
\section{Background}
\label{sec:background}

We use five standard primitives, each recalled where it is applied.
\emph{Amplitude encoding} maps $\mathbf{v}\in\mathbb{R}^N$ to
$|\psi\rangle=\sum_j(v_j/\|\mathbf{v}\|)|j\rangle$, at
$\mathcal{O}(2^{n})$ CNOTs for generic
states~\cite{shende2006synthesis}. \emph{QPE}~\cite{abrams1999quantum,nielsen2010quantum}
is the core subroutine of HHL; first-order \emph{Trotter--Suzuki
Hamiltonian simulation}~\cite{lloyd1996universal,suzuki1990fractal}
drives its controlled-$U$ blocks and the Paradigm-D constructions.
\emph{QSVT} applies a QSP-derived polynomial to the singular values of
a block-encoded
matrix~\cite{gilyen2019quantum,martyn2021grand,low2017optimal}.
\emph{McLachlan's variational
principle}~\cite{mclachlan1964variational,yuan2019theory} yields
$M(\theta)\dot\theta=-V(\theta)$, driving var-QITE and AVQDS.
\emph{Schr\"{o}dingerisation} lifts the non-unitary heat semigroup to a
unitary on an auxiliary momentum register, recovering dissipation by
post-selecting
$p>0$~\cite{jin2024prl,jin2024schrodingerisation,jin2023schrodingerization}.

\section{Problem Formulation and Setup}
\label{sec:problem}

We solve the 1-D heat equation~\eqref{eq:heat} on $x\in(0,L)$,
$L=1$, $u(0,t)=u(L,t)=0$, with diffusivity $\alpha=0.01$.
Second-order central differences on $N$ interior points ($\Delta x=L/(N{+}1)$)
yield the semi-discrete system
\begin{equation}
  \frac{d\mathbf{u}}{dt} = -\alpha\,\mathcal{D}\,\mathbf{u},
  \label{eq:semidiscrete}
\end{equation}
where $\mathcal{D}\in\mathbb{R}^{N\times N}$ is the symmetric positive
definite tridiagonal Laplacian (diagonal~$+2/(\Delta x)^2$,
off-diagonal~$-1/(\Delta x)^2$).
The exact solution via eigendecomposition $\mathcal{D}=V\Lambda V^T$ is
\begin{equation}
  \mathbf{u}(t) = V\,\mathrm{diag}(e^{-\alpha\lambda_k t})\,V^T\,\mathbf{u}(0),
  \label{eq:classical_sol}
\end{equation}
computed classically as the reference baseline.

\subsection{Parameters}
Grid sizes $N=2^n$ for $n\in\{4,5,6,7\}$
($N\in\{16,32,64,128\}$), each encoded into $n$ system qubits.
Per-config $\Delta t$ is chosen so that the CFL-like ratio
$r=\alpha\Delta t/(\Delta x)^2\!\approx\!0.4$ is held constant across
grid sizes.
We integrate to a fixed horizon $T{=}1$; with $r$ held fixed, each~$N$
determines $\Delta t$ and $\Delta x$ as above. The $N$~interior
amplitudes represent only $x\in(0,L)$.

\subsection{Boundary conditions}
\label{sec:bcs}
Homogeneous Dirichlet conditions are chosen because they make
$\mathcal{D}$ symmetric positive definite with a closed-form sine
eigenbasis---supplying the exact classical
reference~\eqref{eq:classical_sol}---and because several kernels are
ported in their published Dirichlet form (QSM's DST primitives and
the Schade dilation
angles)~\cite{klappenecker2001discrete,febrianto2025spectral}.
The choice is not neutral across paradigms: Neumann or Robin
conditions perturb only the boundary rows of $\mathcal{D}$, exchanging
the sine basis for cosine or mixed bases---a DCT-type transform for
the transform kernels~\cite{klappenecker2001discrete}, the physical-BC
construction of~\cite{jin2024schrodingerisation} for
Schr\"{o}dingerisation---while the linear-system and Paradigm-C
kernels see only a modified tridiagonal matrix with $\mathcal{O}(1)$
conditioning under fixed~$r$. In the harness the BC enters only
through the discretised operator and the boundary-value
configuration, so alternative conditions are an extension of the
operator stencil rather than a redesign; we report the Dirichlet
instance on which all eleven kernels are simultaneously well-defined.

\subsection{Initial conditions}
Three profiles stress different parts of the Laplacian spectrum,
quantified by the discrete sine coefficients
$\hat{u}_k(0){=}\sum_j u_j(0)\sin(k\pi j/(N{+}1))$:
(i)~a rectangular \emph{pulse} of unit height over the central
$20\%$ of the interior, whose $|\hat{u}_k(0)|$ decays only as $1/k$ for
odd $k$ and therefore excites a broad band of eigenmodes;
(ii)~a \emph{Gaussian} bump with $\sigma{=}L/6$, whose sine
coefficients decay exponentially in $k$ (low-frequency dominated,
$|\hat{u}_k|{<}10^{-4}|\hat{u}_1|$ for $k{\geq}6$); and
(iii)~a \emph{bimodal} profile
$\sin(\pi x/L)+0.5\sin(3\pi x/L)$, scaled to unit peak, supported
essentially on two discrete modes $k{\in}\{1,3\}$---a multi-mode
stress test with no high-frequency content.
These three profiles are chosen as extreme points of sine-spectrum
occupancy, the variable that drives each paradigm's dominant error
term: Trotter, locality-truncation, and momentum-resolution errors
grow with high-$k$ weight (pulse), whereas smooth data isolates
reconstruction and normalisation effects (Gaussian, bimodal;
\S\ref{sec:norm-ablation}).

\subsection{Output model}
\label{sec:output-model}
All kernels encode temperatures into quantum amplitudes:
$|\mathbf{u}\rangle=\mathbf{u}/\|\mathbf{u}\|$.
Reconstructing $N$ encoded amplitudes to fixed per-component
precision~$\epsilon$ is in the regime of (partial) state tomography, with
a total sampling cost that grows with~$N$ and as~$\epsilon\to 0$
exhibits the familiar $1/\epsilon^2$ Monte Carlo
factor~\cite{haah2017sample}. Correspondingly, practical
``quantum speedup'' accounts for PDEs are more credible when the target
is a \emph{low-dimensional} set of
observables of~$\mathbf{u}$ than when the goal is a full
field readout~\cite{aaronson2015read}.

The statevector backend reads amplitudes directly and is the
zero-shot limit of both output models; sampled backends price them
differently. Computational-basis sampling recovers the full field at
$\mathcal{O}(N/\epsilon^{2})$ total shots (signed amplitudes would
additionally require interferometric or amplitude-estimation
circuits; the non-negative temperature fields used here avoid that
overhead). A compact observable $O=\sum_{j}c_{j}P_{j}$ instead needs
a number of measurement settings independent of~$N$, and operators
diagonal in a structured basis (e.g., $\mathcal{D}$ in its sine
eigenbasis) collapse to a single setting at the price of appending
the transform circuit---a depth-versus-settings trade quantified
in~\S\ref{sec:observable-advantage}.

\subsection{Condition number and scaling}
\label{sec:cond-and-scaling}
For the implicit-Euler matrix $A=I+\alpha\Delta t\,\mathcal{D}$, the
eigenvalues are $1+\alpha\Delta t\,\lambda_k(\mathcal{D})$, so
\begin{equation}
\kappa(A) = \frac{1+\alpha\Delta t\,\lambda_{\max}(\mathcal{D})}
                 {1+\alpha\Delta t\,\lambda_{\min}(\mathcal{D})} .
\label{eq:cond}
\end{equation}
For the Dirichlet Laplacian on $N$ interior points with
$\Delta x=L/(N{+}1)$, the extremal eigenvalues are
$\lambda_{\max}(\mathcal{D})=\tfrac{4}{(\Delta x)^2}\sin^2\!\big(\tfrac{N\pi}{2(N+1)}\big)$
and $\lambda_{\min}(\mathcal{D})=\tfrac{4}{(\Delta x)^2}\sin^2\!\big(\tfrac{\pi}{2(N+1)}\big)$.
With $\Delta t=r(\Delta x)^2/\alpha$ (so $\alpha\Delta t/(\Delta x)^2{=}r$
fixed), these give
$\alpha\Delta t\,\lambda_{\max}(\mathcal{D})\!\to\!4r$ and
$\alpha\Delta t\,\lambda_{\min}(\mathcal{D})=r\pi^2/(N{+}1)^2\!\to\!0$ as $N{\to}\infty$, hence
\begin{equation}
\kappa(A) \;\approx\; \frac{1+4r}{1+r\pi^2/(N{+}1)^2}
         \;\xrightarrow[N\to\infty]{}\; 1+4r .
\label{eq:cond-asymptotic}
\end{equation}
For $r{=}0.4$ this limit is $\kappa(A)\!\to\!2.6$, and $\kappa$ is
scale-stable under our fixed CFL ratio. The implicit-Euler system is
therefore well-conditioned across $n{=}4$--$7$, which is favourable for
HHL and VQLS; in practice the dominant error sources for those kernels
are QPE resolution and ansatz expressibility respectively, not $\kappa$.

\section{Kernels Under Test}
\label{sec:kernels}
Within each paradigm, kernels share reconstruction assumptions and
often exhibit similar cost or failure modes; subsections give idea,
circuit mechanism, and accuracy bounds. Backend omissions are
consolidated in \S\ref{sec:backends}.

\subsection{Paradigm A --- Quantum Linear Solvers}
\label{sec:paradigm-a}
Paradigm-A kernels perform \emph{coherent inversion}: they apply a
unitary transformation to a block encoding of $A$ to produce the
solution state $|x\rangle\propto A^{-1}|b\rangle$ directly on the
quantum register, without classical optimisation. All three kernels
discretise~\eqref{eq:heat} in time with implicit Euler,
\begin{equation}
  A\,\mathbf{x} = \mathbf{b},\quad
  A = I + \alpha\Delta t\,\mathcal{D},\quad
  \mathbf{b} = \mathbf{u}^n,\quad
  \mathbf{x} = \mathbf{u}^{n+1},
  \label{eq:linsys}
\end{equation}
and differ in how they realise the inversion step.

\textbf{HHL} is the canonical quantum
linear-system algorithm, recovering
$|x\rangle\propto A^{-1}|b\rangle$ through eigenvalue inversion.
It applies QPE with $m{=}6$ clock qubits to extract $\tilde\lambda_j$,
a controlled ancilla rotation $C/\tilde\lambda_j$ to encode the
inverse, and post-selects the ancilla to recover the solution ~\cite{harrow2009quantum}. The
QPE block dominates the circuit cost through repeated
Hamiltonian-simulation--based controlled-$e^{iAt}$ unitaries; the
overall depth depends on the cost of simulating $A$ and the chosen
implementation. Accuracy is bounded by the $2^{-m}$ eigenvalue
resolution, which sets the ancilla rotation angles. Among
Paradigm-A kernels, HHL represents the eigenvalue-by-eigenvalue
inversion strategy that QSVT and QLS-Fourier replace with polynomial
and Fourier alternatives.
On the \emph{noisy} Aer path, transpiling the deep controlled
Hamiltonian-simulation/QPE stack to native two-qubit gates triggers
Qiskit two-qubit unitary decomposition failures for
$n\in\{5,6,7\}$ (we observe this as a
\texttt{TwoQubitWeylDecomposition} error during synthesis); we
therefore include \textbf{HHL} on the noisy backend only at $n{=}4$.
Statevector and ideal-shot runs remain for $n{=}4$--$7$.

\textbf{QSVT} applies a polynomial approximation of \(1/x\) to the singular values of a block-encoded matrix. \(U_A\) block-encodes \(\tilde{A}=A/\|A\|_2\), and \(d+1\) alternating \(U_A/U_A^\dagger\) queries interleaved with QSP phase rotations realise the polynomial at degree \(d=15\). The approximation is taken over the normalized singular-value interval \([a,1]\), where \(a=\sigma_{\min}(\tilde{A})\approx 1/\kappa(A)\). Oracle-call count is constant in \(n\); circuit depth depends on the block-encoding cost. Accuracy is bounded by the degree-\(d\) polynomial approximation error on \([a,1]\), which decays as $\mathcal{O}((1{-}a)^d)$ for well-conditioned $A$.~\cite{gilyen2019quantum,martyn2021grand} .

\textbf{QLS-Fourier} denotes our Fourier/LCU-style functional quantum
linear-system implementation. We use this label for the benchmark kernel
based on the linear-system framework of~\cite{childs2017quantum} and the
functional-QLS comparison setting of~\cite{lefterovici2026functional}.
It approximates $1/x$ by a truncated Fourier integral implemented as an
LCU of Hamiltonian-simulation steps. We evaluate the post-selected
LCU in the DST eigenbasis, where $\tilde{A}$ is diagonal, reducing
the cost to a scalar evaluation per eigenmode. Accuracy is bounded
by the Fourier truncation and discretisation schedule
of~\cite{lefterovici2026functional}.

The Appendix-D schedule of~\cite{lefterovici2026functional} yields
$J\!\cdot\!2L\approx 10^{6}$ LCU terms at $n{=}4$, growing to
${\sim}9\times10^{7}$ ($n{=}6$) and ${\sim}1.6\times10^{9}$ ($n{=}7$);
compiling a controlled $e^{i\tilde A t_j}$ per term is infeasible, and
truncating to a tractable subset substitutes an ad-hoc cutoff for the
algorithm's error budget. We instead evaluate the ancilla-$|0\rangle$
block of PREPARE--SELECT--PREPARE$^{\dagger}$ analytically in the
eigenbasis of $\tilde A$, equivalent to a noiseless statevector run of
the full LCU circuit. As this path bypasses both transpilation and
sampling, QLS-Fourier is evaluated statevector-only
(\S\ref{sec:backends}).

\subsection{Paradigm B --- NISQ Variational Solver}
\label{sec:paradigm-b}

\textbf{VQLS} leverages shallow, parameterized circuits suitable for current NISQ devices\cite{lu2026distributedvqls,bravo2023variational}. It targets the same implicit-Euler system~\eqref{eq:linsys} by minimising a local
Hadamard-test cost function whose global minimum coincides with
$|x\rangle\propto A^{-1}|b\rangle$; a hardware-efficient ansatz whose
depth adapts to the register (four--five layers at the tested sizes,
matching Table~\ref{tab:circuit_specs}) is optimised classically and
warm-started across time steps.
In the Pauli basis the tridiagonal $A$ is \emph{not} sparse: its
exact Hilbert--Schmidt expansion spreads over $L{=}N{=}2^{n}$ Pauli
strings (verified numerically for $n{=}3$--$7$);
$\mathcal{O}(n)$-term decompositions are known only over non-Pauli
primitives such as $\sigma^{\pm}$~\cite{liu2021poisson}. Our
implementation truncates the expansion at $10\%$ of the leading
coefficient, retaining $L{=}4$ dominant strings ($I$, $X_{0}$,
$X_{0}X_{1}$, $Y_{0}Y_{1}$) at every tested $n$ at the cost of a
${\sim}15\%$ spectral-norm perturbation of $A$---a systematic bias
shared by all VQLS results reported here. Accuracy is bounded by
this truncation, by ansatz expressibility, and by classical-optimiser
convergence in the non-convex landscape. VQLS needs no
fault-tolerant resources, but its sampled cost is multiplicative
($L^{2}$ Hadamard tests per cost evaluation $\times$
$\mathcal{O}(10^{2})$ optimiser iterations $\times$ time steps), so
we run it statevector-only (\S\ref{sec:backends}).

\subsection{Paradigm C --- Imaginary-Time Variational Dynamics}
\label{sec:paradigm-c}
Paradigm-C kernels approximate the heat semigroup
$e^{-\alpha\mathcal{D}t}$ directly in imaginary time via a variational
Pauli basis, bypassing both the linear system and any unitary
dilation. The var-QITE and AVQDS kernels follow the constructions of
Alipanah~et~al.~\cite{alipanah2025avqds}.

\textbf{QITE} realises each imaginary-time
step through a best-unitary approximation of $e^{-H\Delta\tau}$
expressed in a local Pauli basis ~\cite{motta2020determining}. At each step, Pauli expectations
$\langle\sigma_j\rangle$ and $\langle\sigma_j H\rangle$ are measured,
coefficients $a_j$ are obtained by solving a Gram-matrix linear
system, and the state is updated via a first-order Trotter product
$\prod_j e^{-i\Delta\tau\,a_j\sigma_j}$. The benchmark uses compressed
QITE with locality support $D=3$, chosen as the empirical
accuracy--runtime sweet spot; the classical inner loop requires
$\mathcal{O}(N_\sigma^2)$ expectation evaluations per step, where
$N_\sigma$ is the number of retained Pauli terms in the compressed
$\sigma$ basis. Accuracy is bounded by the truncation support $D$
and by the Trotter commutator error
$\mathcal{O}(\|a\|^2\Delta\tau^2)$. QITE is the reference
imaginary-time method against which the variational relaxations
var-QITE and AVQDS are measured.

\textbf{Var-QITE} is a
variational relaxation of QITE that replaces the exact statevector
with a parameterised ansatz ~\cite{mclachlan1964variational,yuan2019theory}. Parameters evolve under McLachlan's
principle $M(\theta)\dot\theta{=}-V(\theta)$, trading QITE's
$\mathcal{O}(4^n)$ observable set for a compact $p{\times}p$ gradient
system; accuracy is bounded by ansatz expressibility.

\textbf{AVQDS} is an adaptive extension of
var-QITE in which the ansatz is grown online ~\cite{alipanah2025avqds}. A new Pauli rotation
drawn from a classical operator pool is appended whenever the
McLachlan residual exceeds a threshold; circuit depth is determined
during simulation rather than in advance. Accuracy is bounded by the
residual threshold and operator-pool coverage.

\subsection{Paradigm D --- Real-Time Hamiltonian Simulation /
Unitary Dilation}
\label{sec:paradigm-d}
Paradigm-D kernels handle the heat equation's non-unitary dynamics
through real-time evolution, either via a classical scalar correction
(Hamiltonian simulation) or by embedding the dissipative dynamics
into a unitary on an enlarged Hilbert space (Schade-Hamiltonian,
Schr\"{o}dingerisation).

\textbf{Hamiltonian simulation} is the
simplest real-time approach: $e^{-iH\Delta t}$ with
$H{=}\alpha\mathcal{D}$ is applied via Lie--Trotter and the
dissipative decay is recovered classically---the real part of the
evolved state is rescaled by $\|T\|\,e^{-\langle H\rangle\Delta t}$
using the pre-step energy ~\cite{alipanah2025avqds}. Accuracy is bounded by the mean-field
approximation: every mode relaxes at $\langle H\rangle$ at leading
order in $\Delta t$, with mode-specific corrections only at
$\mathcal{O}(\Delta t^2)$. No auxiliary register, no post-selection.

\textbf{Schade-Hamiltonian} is a
unitary-dilation approach that embeds the heat contraction
$e^{-\alpha\mathcal{D}\Delta t}$ as the $|0\rangle_{\mathrm{anc}}$
block of a real-time unitary on an ancilla-extended register. We port
the Cholesky/block-off-diagonal embedding
of~\cite{schade2023hamiltoniansim} from the elastic wave setting to
the dissipative heat setting via exact per-step dilation angles
$\theta_k{=}\arccos(e^{-\alpha\lambda_k\Delta t})$ computed in the
Laplacian eigenbasis; $U{=}e^{-iH}$ is applied as a dense unitary and
post-selected on the ancilla. Within this benchmark this is therefore a \emph{structured dilation
reference}: it tests the effect of an eigenbasis-aligned unitary
embedding rather than claiming a sparse, hardware-optimal construction.
Accuracy is limited by floating-point arithmetic in the eigendecomposition
and by the stated post-selection/readout model, and per-mode decay is
restored exactly.

\textbf{Schr\"{o}dingerisation
(SZ)}
is the second unitary-dilation kernel. Instead of a single ancilla,
SZ appends an $n_p$-qubit momentum register ($n_p{=}n$) on which the
heat equation lifts to a conservative advection equation;
post-selecting on $p{>}0$ recovers the dissipative dynamics ~\cite{jin2024prl,jin2024schrodingerisation,jin2023schrodingerization}.
The compiled DiagonalGate yields $\mathcal{O}(2^n)$ two-qubit gates,
so SZ is statevector-only (\S\ref{sec:backends}). Accuracy is bounded
by momentum-register resolution and the post-selection cut-off.

\subsection{Paradigm E --- Spectral / Transform-Based}
\label{sec:paradigm-e}

\textbf{QSM (Spectral)}
implements the exact propagator~\eqref{eq:classical_sol} as a
three-stage circuit: forward DST into the Laplacian eigenbasis;
multiplexed $R_y$ encoding the eigenmode decay
$|k\rangle|0\rangle\!\to\!e^{-\alpha\lambda_k t}|k\rangle|0\rangle{+}\sqrt{1{-}e^{-2\alpha\lambda_k t}}\,|k\rangle|1\rangle$; inverse DST ~\cite{klappenecker2001discrete,febrianto2025spectral}. Compiled via the QFT-based multiplexed-$R_y$
decomposition, the two-qubit cost scales as $\mathcal{O}(n^2)$ in the
structured implementation we count. The decay is encoded using one
work qubit; extracting the contracted heat state therefore requires
conditioning on the work-qubit outcome or, for observables, incorporating
the success probability into the estimator. QSM serves as the benchmark's
structural reference for per-mode exact decay rather than as a claim of
near-term end-to-end advantage.

\section{Experimental Setup}
\label{sec:experimental}

\subsection{Computational Environment}
All experiments ran on a 48-core Intel Xeon Platinum~8260M workstation
(two sockets, 2.40\,GHz) with 186\,GB RAM, CentOS~8, Python~3.10. The
software stack uses Qiskit~2.3+\cite{qiskit2024} with
Qiskit-Aer~0.17+ for shot-based and noisy simulation, Qiskit IBM
Runtime~0.45+ for the real-hardware interface, and the standard
NumPy~1.24+ / SciPy~1.10+ / Matplotlib~3.10+ scientific stack.
The $n{=}4$--$7$ range is set by the full factorial design rather
than simulator capability: at $n{=}7$, noisy-simulator runs already
take on the order of days for several kernels
(\S\ref{sec:comparison}), while statevector-only studies of
individual kernels can go higher within the same harness.

\subsection{Simulation Backends}
\label{sec:backends}

The framework implements four \texttt{BaseBackend} classes:
statevector, ideal shot-based simulation, noisy simulation, and real
hardware. We report the first three; the hardware interface is
implemented but left for future work.
Table~\ref{tab:backend_compact} summarises the evidence level
associated with each backend. A dash in the results tables denotes
that a run was not performed for the stated implementation reason, not
that the method has zero error.

\begin{table}[htbp]
{\renewcommand{\arraystretch}{1.0}%
\caption{Backend evidence levels used in the benchmark.}
\label{tab:backend_compact}
\centering
\footnotesize
\setlength{\tabcolsep}{4pt}
\begin{tabular}{p{0.16\columnwidth} p{0.76\columnwidth}}
\toprule
Backend & Role and limitations \\
\midrule
SV &
Direct double-precision amplitude evolution. Isolates intrinsic
algorithmic and reconstruction error (Trotter truncation, polynomial
degree, ansatz expressibility, Fourier truncation, dilation error).
The SV columns of
Tables~\ref{tab:summary-n4}--\ref{tab:summary-n7} are produced here.
\\

Id &
Noiseless \texttt{AerSimulator} sampling at
$N_{\mathrm{shots}}{=}100{,}000$ per step (sampling floor
$\sim 1/\sqrt{N_{\mathrm{shots}}}\approx 3\times10^{-3}$). For
variational/dynamics kernels, only expectation estimates are sampled;
classical solves and reconstruction remain classical. \\

Ns &
Same transpiled circuits as Id (not retranspiled, so accuracy
differences are attributable to noise alone), with a portable
gate/readout model: $10^{-3}$ single-qubit depolarising,
$10^{-2}$ two-qubit depolarising, and $10^{-2}$ symmetric readout
bit-flip. $T_1/T_2$ and coupling-map constraints are omitted. \\

HW &
IBM~Runtime \texttt{Sampler}/\texttt{Estimator} via
\texttt{create\_backend("hardware")}; results are reserved for future
work. \\
\bottomrule
\end{tabular}}
\end{table}

The Id backend omits \textbf{VQLS}, \textbf{Schr\"{o}dingerisation},
and \textbf{QLS-Fourier} for the implementation reasons given in
\S\ref{sec:paradigm-b}, \S\ref{sec:paradigm-d},
and~\S\ref{sec:paradigm-a}. The noisy backend uses the same
shot-compatible kernel set, except that \textbf{HHL} is included only
at $n{=}4$ because larger noisy HHL circuits fail during two-qubit
synthesis.

\section{Comparative Evaluation and Results}
\label{sec:comparison}

We present results across all three backends and initial conditions
(Figs.~\ref{fig:sv_results}--\ref{fig:noisy_results}).
Table~\ref{tab:circuit_specs} gives native depth and 1Q/2Q counts by
kernel and~$n$; Tables~\ref{tab:summary-n4} and~\ref{tab:summary-n7}
consolidate terminal relative $\ell_2$ error at $n{=}4$ and $n{=}7$
respectively, for every initial condition and backend column.
Although several kernels take on the order of days to complete a
single $n{=}7$ run with the noisy simulator, every kernel admitted by
each backend finished, and the figures display the full
$n{=}4$--$7$ range.

\subsection{Time Evolution and $\ell_2$ Error Plots}

Each backend figure (Figs.~\ref{fig:sv_results}--\ref{fig:noisy_results})
plots temperature $T(x)$ at five milestones per IC and grid size (top
row) and relative $\ell_2$ error vs.\ simulation progress (bottom row).
Table~\ref{tab:summary-n7}'s SV columns match the $n{=}7$ terminal values
of the bottom-row curves.

\subsubsection{Statevector Backend}
\begin{figure*}[htbp]
\centering
\begin{subfigure}[b]{0.32\textwidth}
  \centering
  \includegraphics[width=\textwidth]{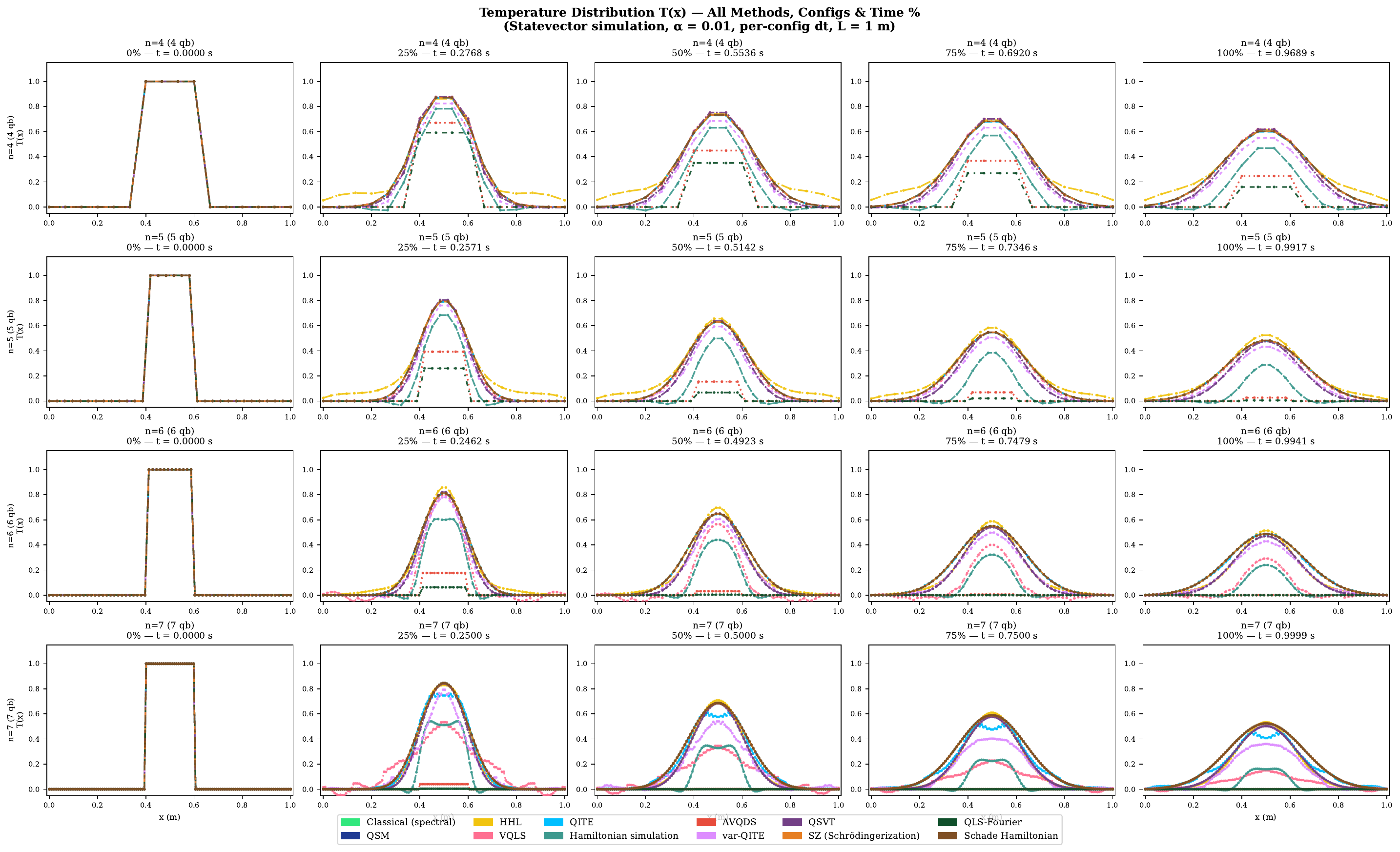}
  \caption{Pulse --- time evolution}
\end{subfigure}\hfill
\begin{subfigure}[b]{0.32\textwidth}
  \centering
  \includegraphics[width=\textwidth]{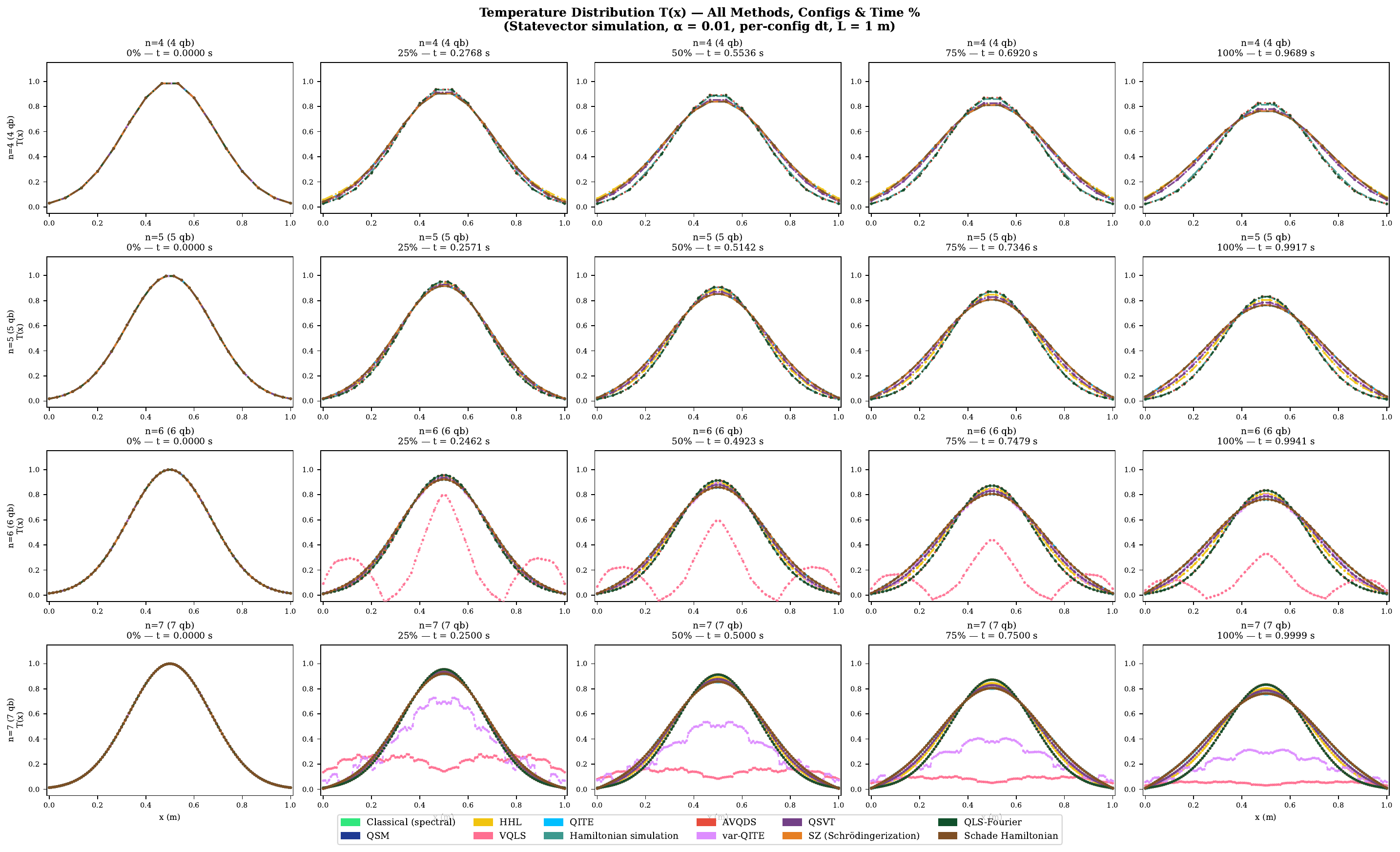}
  \caption{Gaussian --- time evolution}
\end{subfigure}\hfill
\begin{subfigure}[b]{0.32\textwidth}
  \centering
  \includegraphics[width=\textwidth]{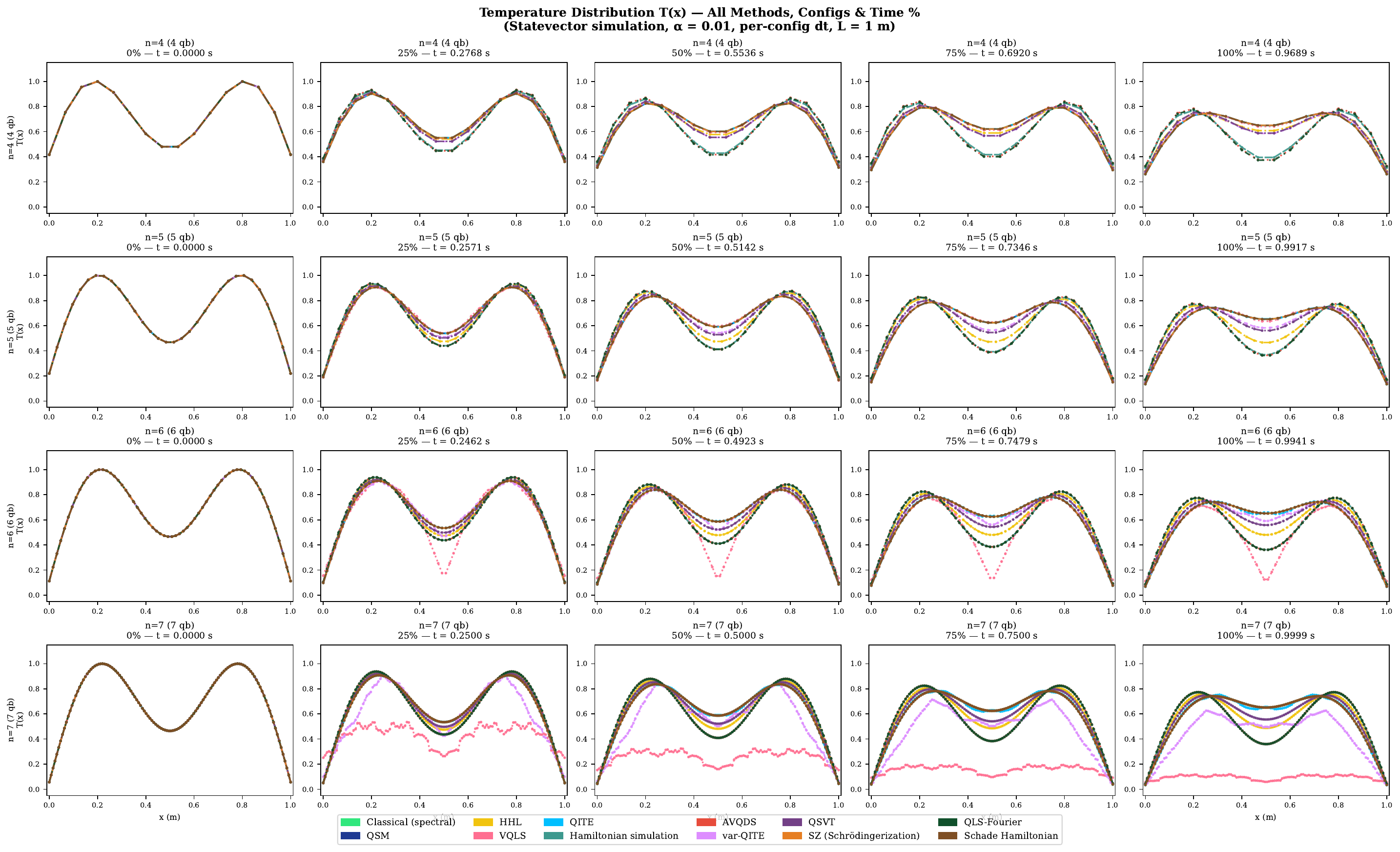}
  \caption{Bimodal --- time evolution}
\end{subfigure}\\[4pt]
\begin{subfigure}[b]{0.32\textwidth}
  \centering
  \includegraphics[width=\textwidth]{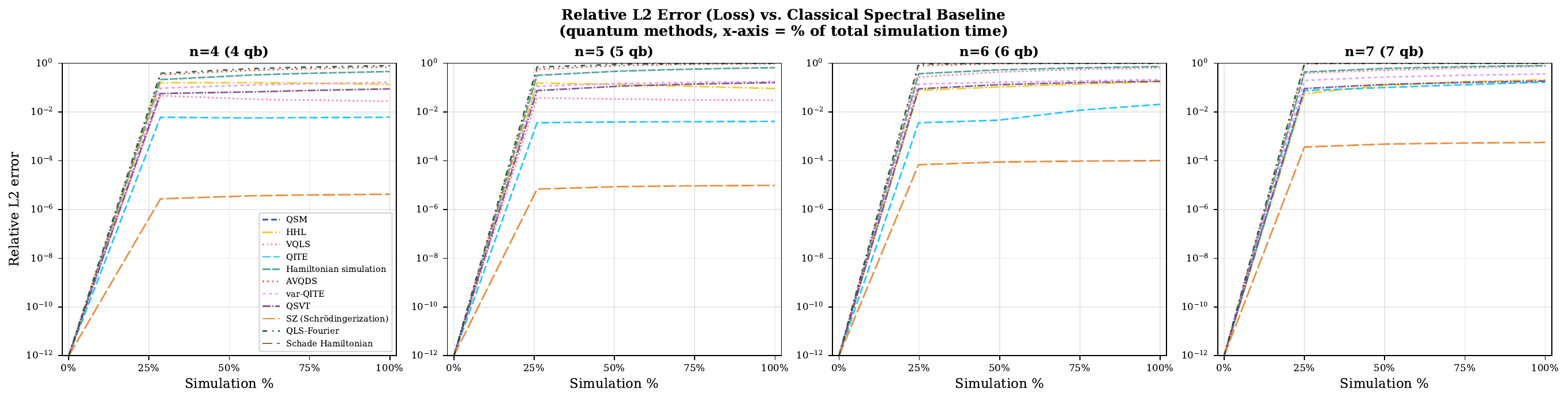}
  \caption{Pulse --- $\ell_2$ error}
\end{subfigure}\hfill
\begin{subfigure}[b]{0.32\textwidth}
  \centering
  \includegraphics[width=\textwidth]{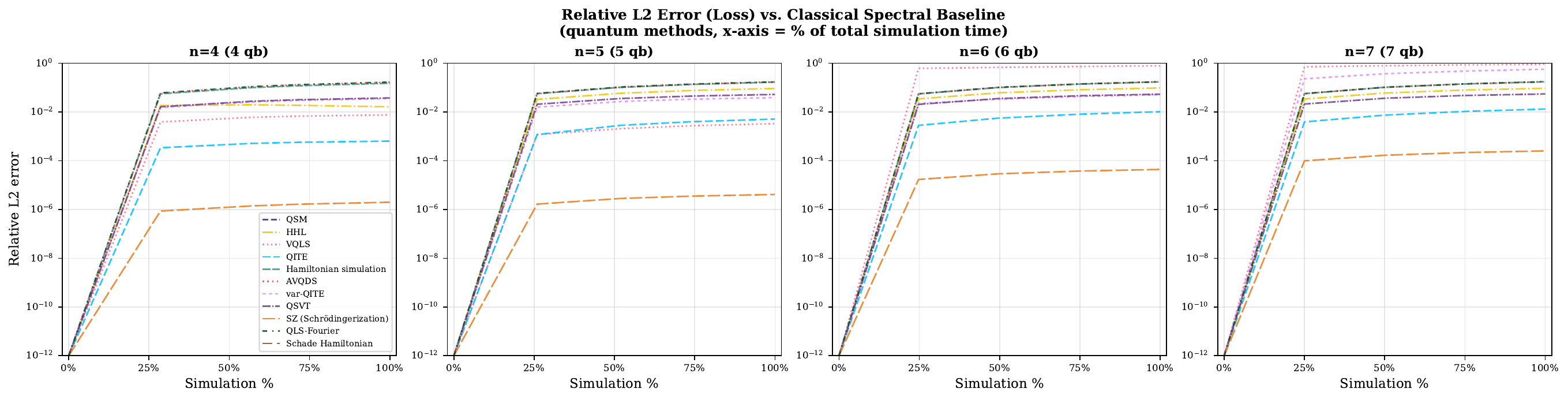}
  \caption{Gaussian --- $\ell_2$ error}
\end{subfigure}\hfill
\begin{subfigure}[b]{0.32\textwidth}
  \centering
  \includegraphics[width=\textwidth]{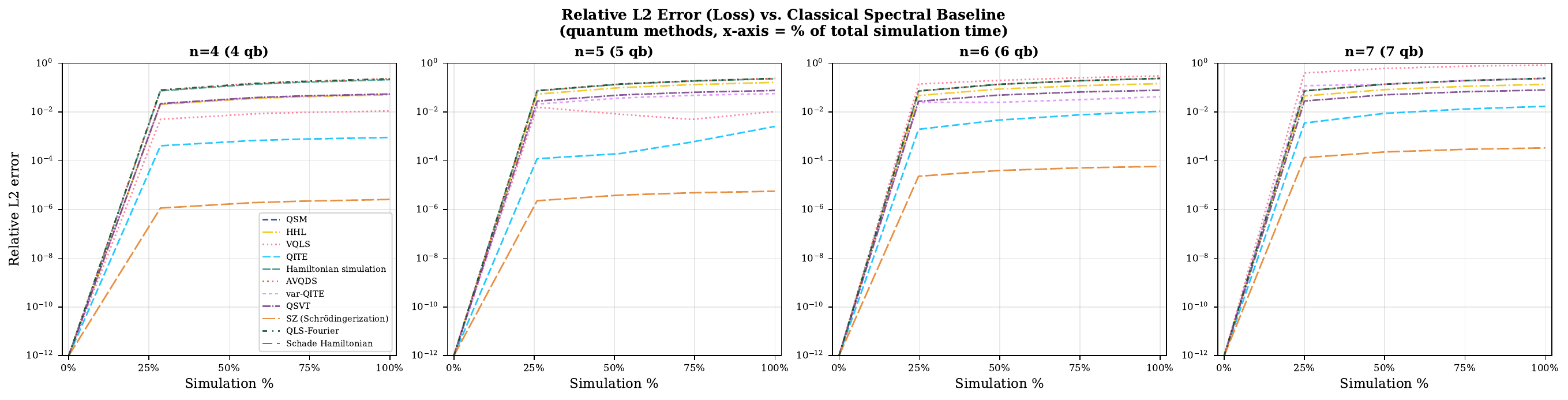}
  \caption{Bimodal --- $\ell_2$ error}
\end{subfigure}
\caption{\textbf{Statevector backend.}
  \textbf{Top row:} temperature evolution $T(x)$ at five time milestones
  across grid sizes $n{=}4$--$7$, one panel per initial condition.
  \textbf{Bottom row:} relative $\ell_2$ error vs.\ simulation progress (\%).
  In each panel, the eleven quantum kernels appear as separate colored
  curves. kernel-level behaviour is discussed in \S\ref{sec:comparison}.}
\label{fig:sv_results}
\end{figure*}

Figure~\ref{fig:sv_results} provides the clearest view of intrinsic
algorithmic behavior because it removes both sampling noise and device
noise. 

The statevector results show a clear hierarchy. QSM and
Schade-Hamiltonian are the most accurate methods overall: across the
three initial conditions---high-frequency (pulse), smooth (Gaussian), and
smooth multi-mode (bimodal)---they overlap the classical baseline to
floating-point precision and stay on the error floor
throughout the run (Table~\ref{tab:summary-n7}, SV). Their success is
structural: both use a basis aligned with the Laplacian eigenstructure.

Schr\"{o}dingerisation is the next most accurate method, but it no longer sits on the
machine-precision floor. It remains very close to the reference,
indicating that the unitary-dilation route captures the heat dynamics
well, but its more general embedding leaves a visible residual gap to the
spectral reference methods. QITE is the most accurate non-transform kernel.
It performs best on the smooth Gaussian and bimodal benchmarks, and
degrades on the high-frequency pulse as $n$ increases (local
truncation and high-frequency content).

The middle-to-lower tier: HHL and QSVT are inversion-limited (phase
estimation vs.\ polynomial degree); their 2Q counts in
Table~\ref{tab:circuit_specs} grow fastest among included kernels. VQLS
fades with~$n$ (capped-depth ansatz); var-QITE stays below QITE
(Table~\ref{tab:summary-n7}, SV) on the reduced variational family.

AVQDS, Hamiltonian simulation, and QLS-Fourier share a smooth-profile
error \emph{band} (Table~\ref{tab:summary-n7}, SV Gaussian/bimodal); on
pulse, AVQDS and QLS-Fourier saturate while Hamiltonian simulation stays
lower but still large. QLS-Fourier in Table~\ref{tab:circuit_specs} is
listed only for $n{=}4$ (decomposition cap), independent of that failure
mode.

\subsubsection{Ideal (Shot-Based) Backend }

\begin{figure*}[htbp]
\centering
\begin{subfigure}[b]{0.32\textwidth}
  \centering
  \includegraphics[width=\textwidth]{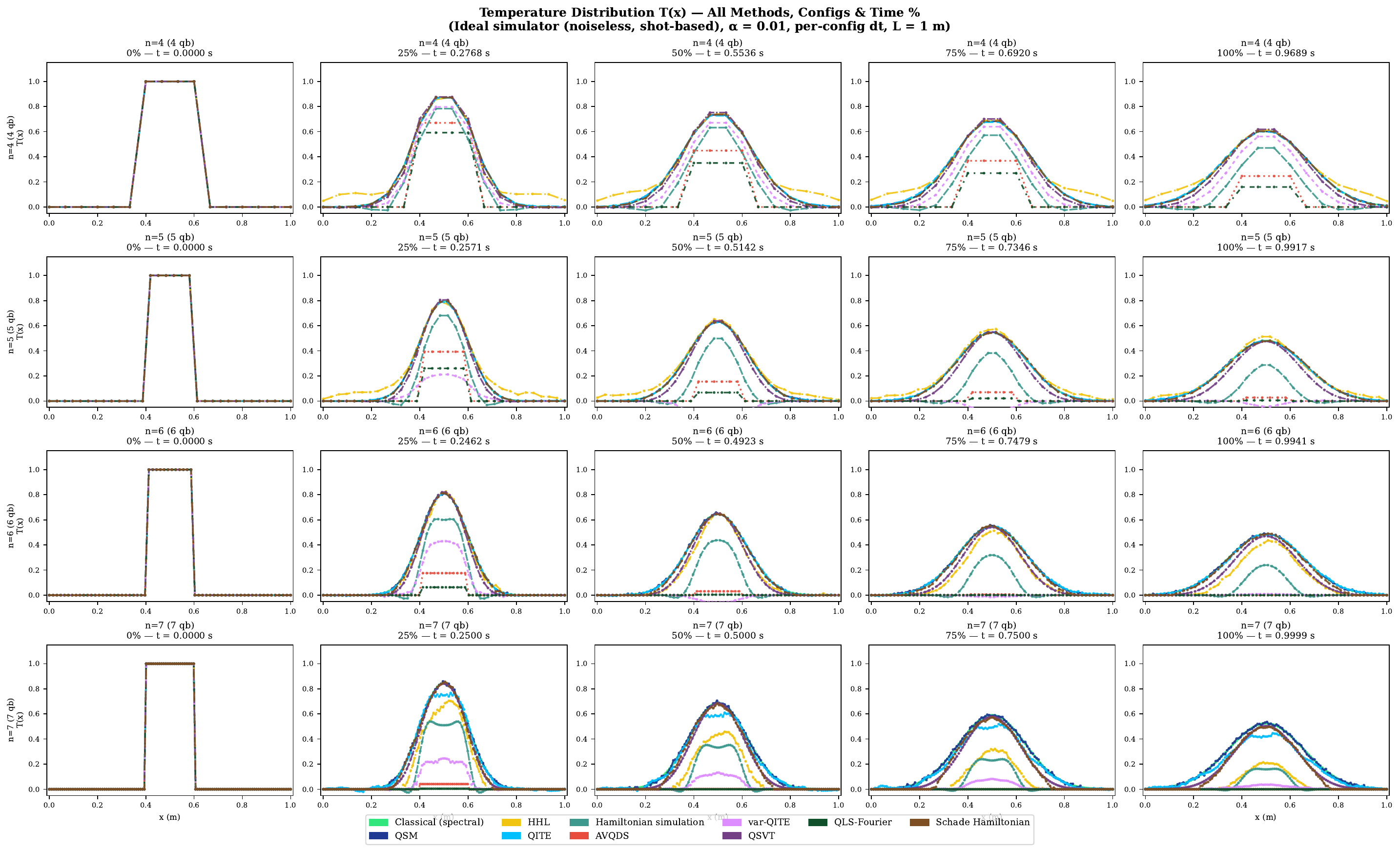}
  \caption{Pulse --- time evolution}
\end{subfigure}\hfill
\begin{subfigure}[b]{0.32\textwidth}
  \centering
  \includegraphics[width=\textwidth]{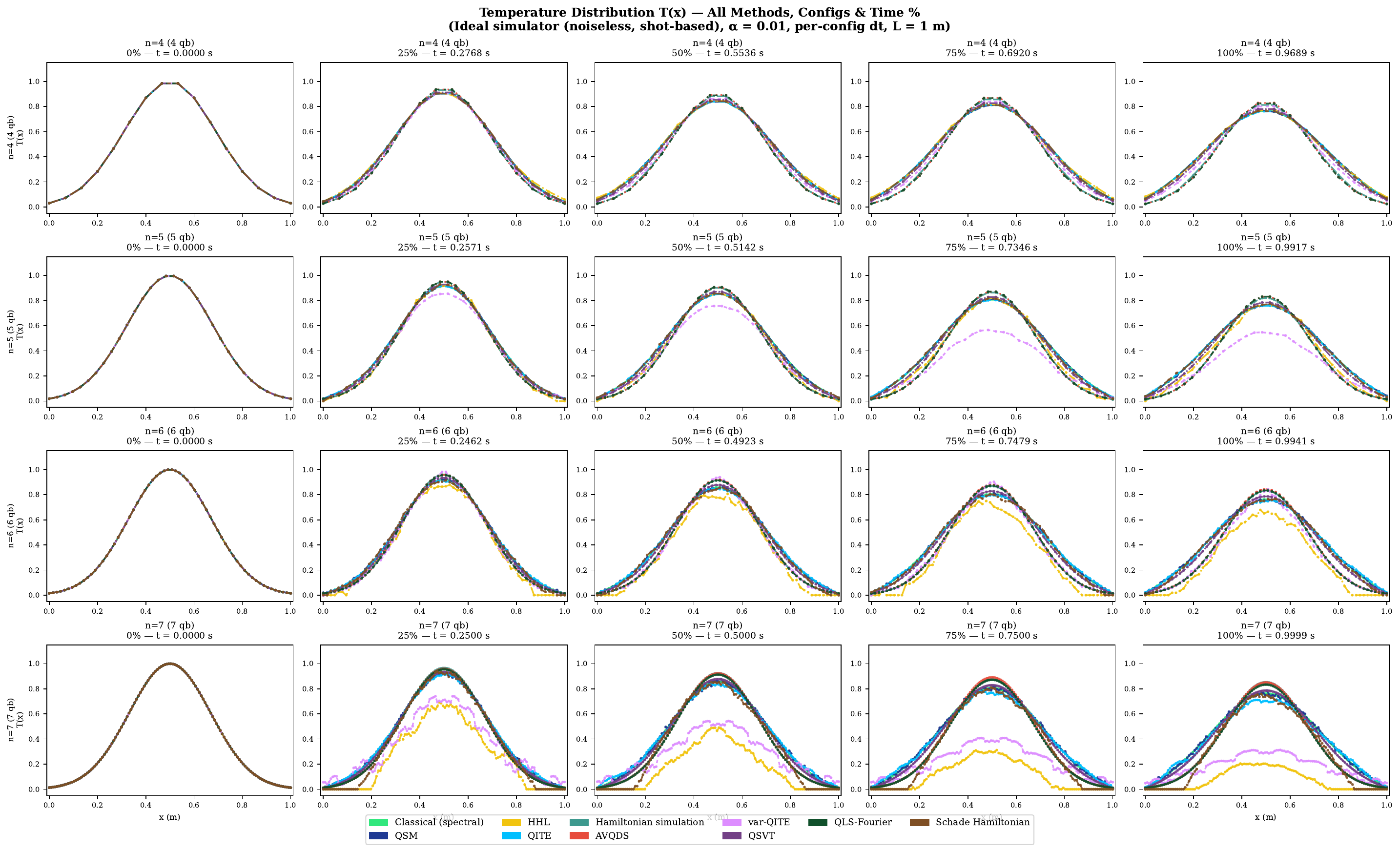}
  \caption{Gaussian --- time evolution}
\end{subfigure}\hfill
\begin{subfigure}[b]{0.32\textwidth}
  \centering
  \includegraphics[width=\textwidth]{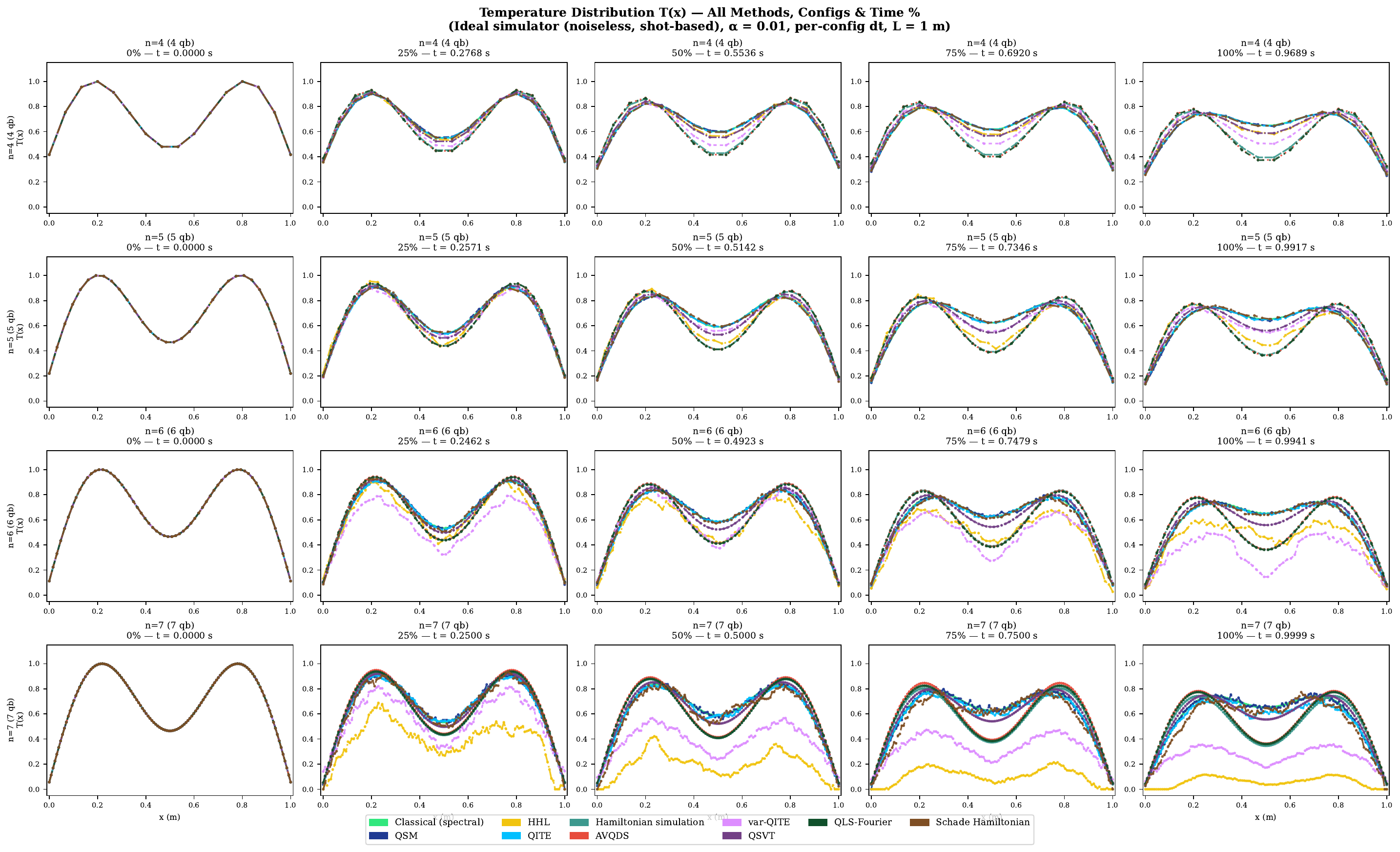}
  \caption{Bimodal --- time evolution}
\end{subfigure}\\[4pt]
\begin{subfigure}[b]{0.32\textwidth}
  \centering
  \includegraphics[width=\textwidth]{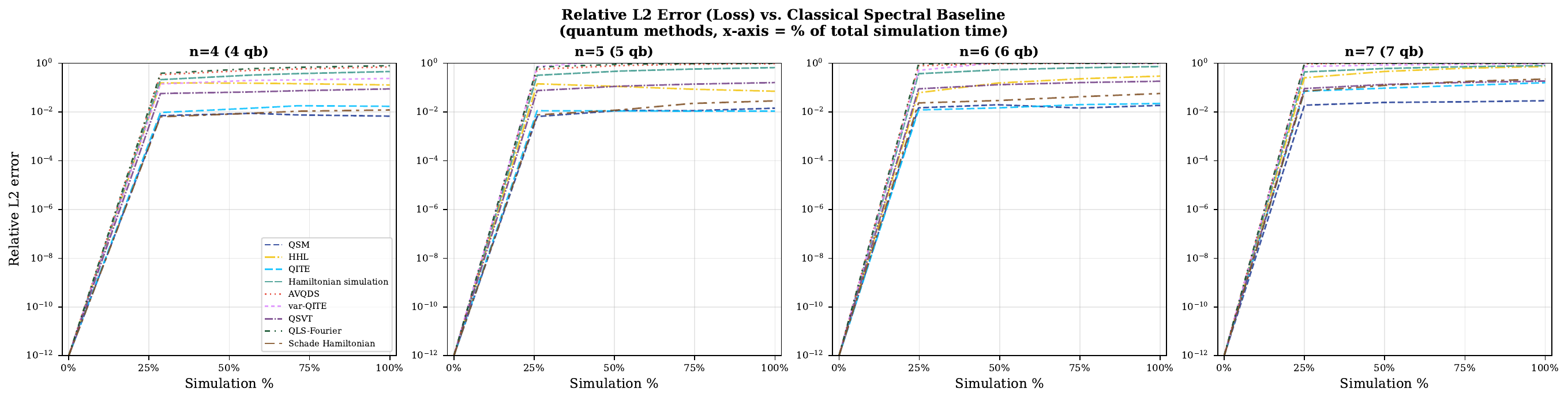}
  \caption{Pulse --- $\ell_2$ error}
\end{subfigure}\hfill
\begin{subfigure}[b]{0.32\textwidth}
  \centering
  \includegraphics[width=\textwidth]{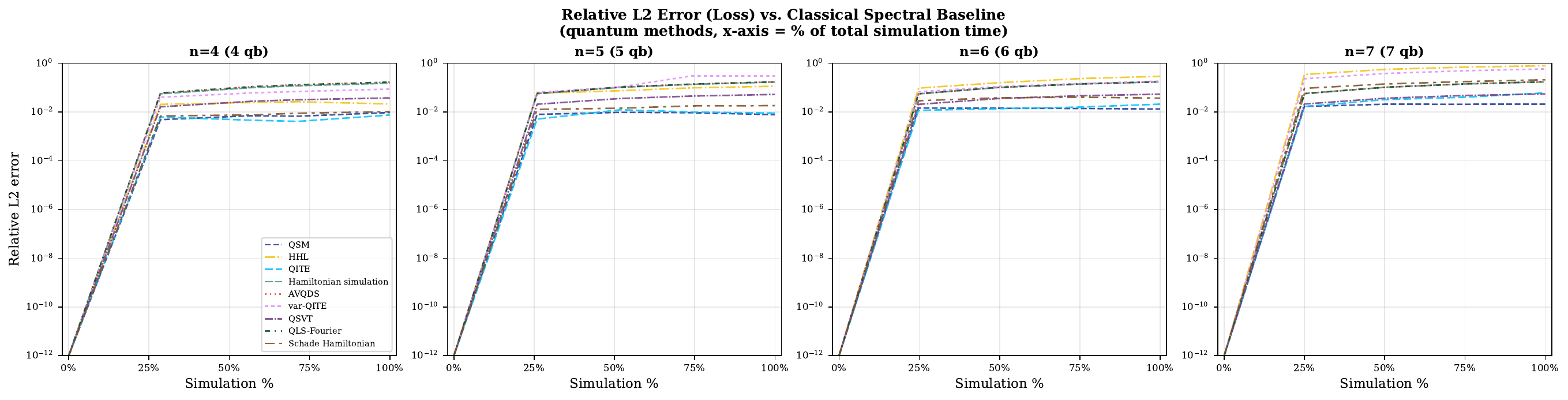}
  \caption{Gaussian --- $\ell_2$ error}
\end{subfigure}\hfill
\begin{subfigure}[b]{0.32\textwidth}
  \centering
  \includegraphics[width=\textwidth]{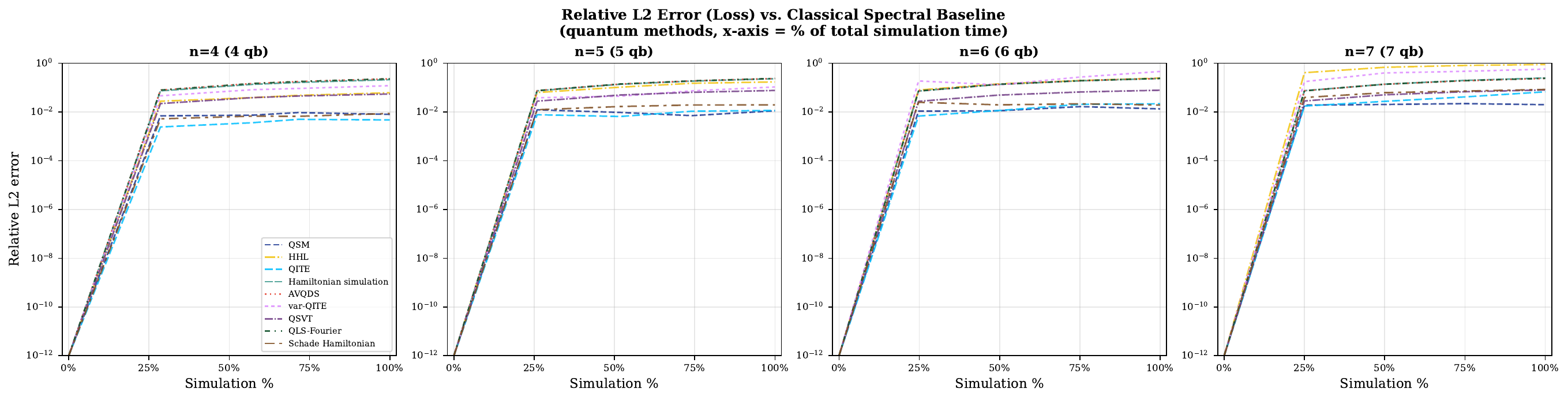}
  \caption{Bimodal --- $\ell_2$ error}
\end{subfigure}
\caption{\textbf{Ideal (shot-based, noiseless) backend}
  ($N_{\mathrm{shots}}{=}100\,000$).
  VQLS, Schr\"{o}dingerisation, and QLS-Fourier are omitted
  (statevector-only; \S\ref{sec:backends}).}
\label{fig:ideal_results}
\end{figure*}

Figure~\ref{fig:ideal_results} keeps the broad statevector ranking under
finite shots; the floor near ${\sim}1/\sqrt{N_{\mathrm{shots}}}$ caps how
low errors can go once sampling dominates. The ideal (Id) columns in
Table~\ref{tab:summary-n7} match the plateaus (e.g.\ QSM and
Schade-Hamiltonian in the $10^{-2}$--$10^{-1}$ range, and the shared
Gaussian plateau for Hamiltonian simulation and AVQDS; QLS-Fourier
shares the same SV plateau but is not sampled here for the reasons
in \S\ref{sec:paradigm-a}). QSM and Schade-Hamiltonian still hug the
reference on smooth data; QITE is still the strongest non-transform
kernel. The pulse separates methods more
clearly in the time-evolution and error panels as~$n$ increases. HHL/QSVT
and the middle group keep the same ordering as on SV, compressed by
statistics rather than reordered.

\subsubsection{Noisy Simulator Backend}

\begin{figure*}[htbp]
\centering
\begin{subfigure}[b]{0.32\textwidth}
   \centering
  \includegraphics[width=\textwidth]{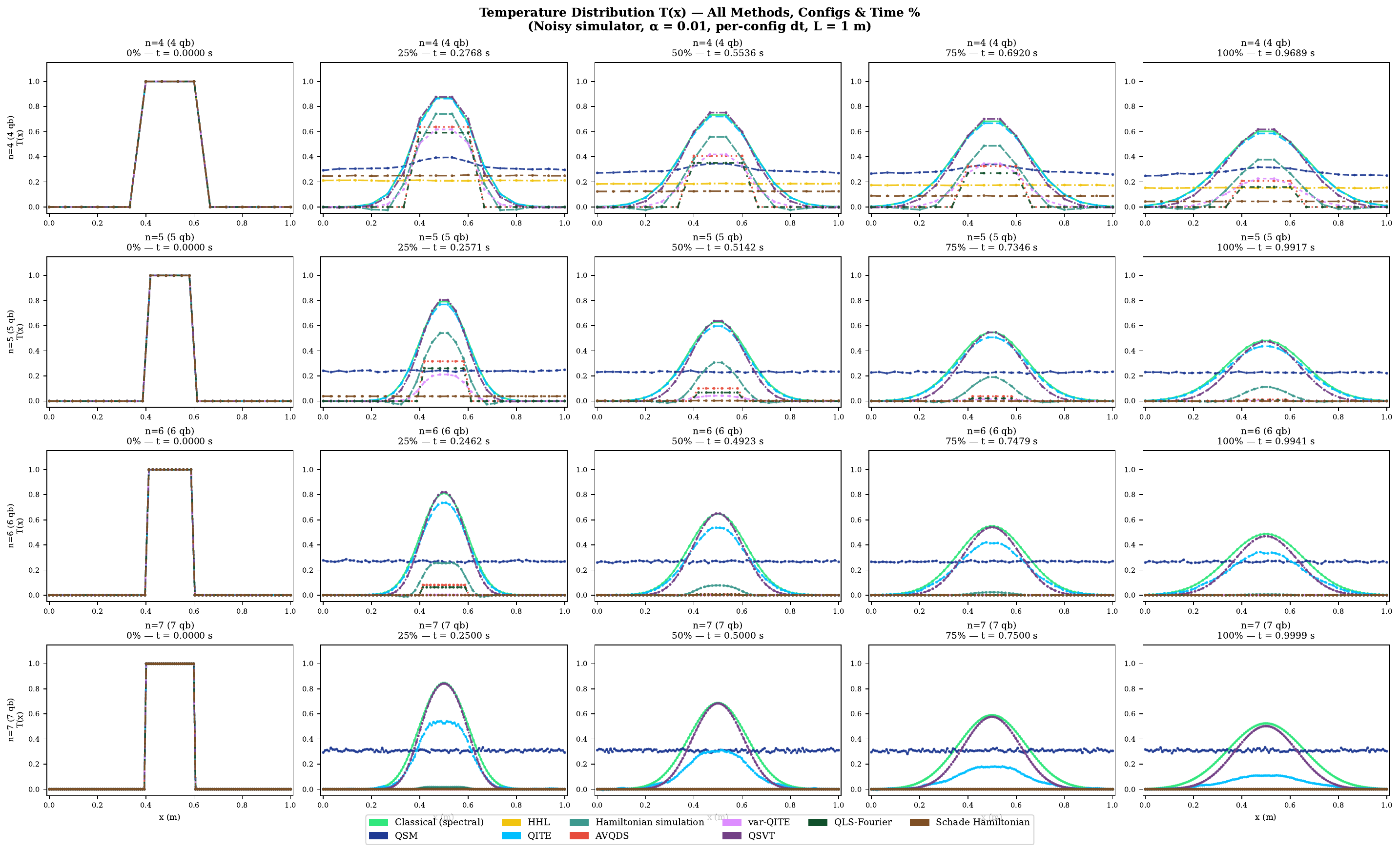}
  \caption{Pulse --- time evolution}
\end{subfigure}\hfill
\begin{subfigure}[b]{0.32\textwidth}
  \centering
  \includegraphics[width=\textwidth]{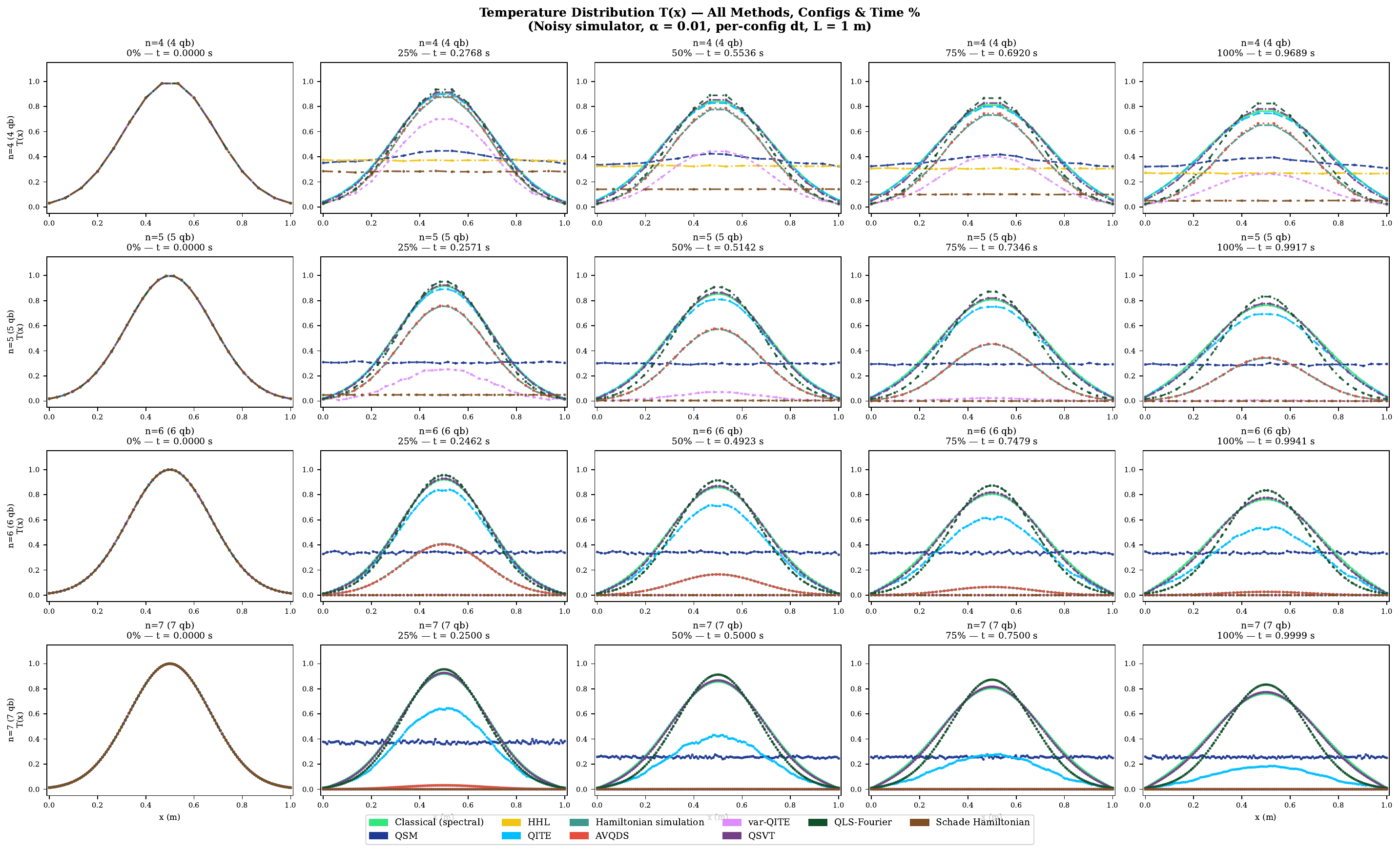}
  \caption{Gaussian --- time evolution}
\end{subfigure}\hfill
\begin{subfigure}[b]{0.32\textwidth}
  \centering
  \includegraphics[width=\textwidth]{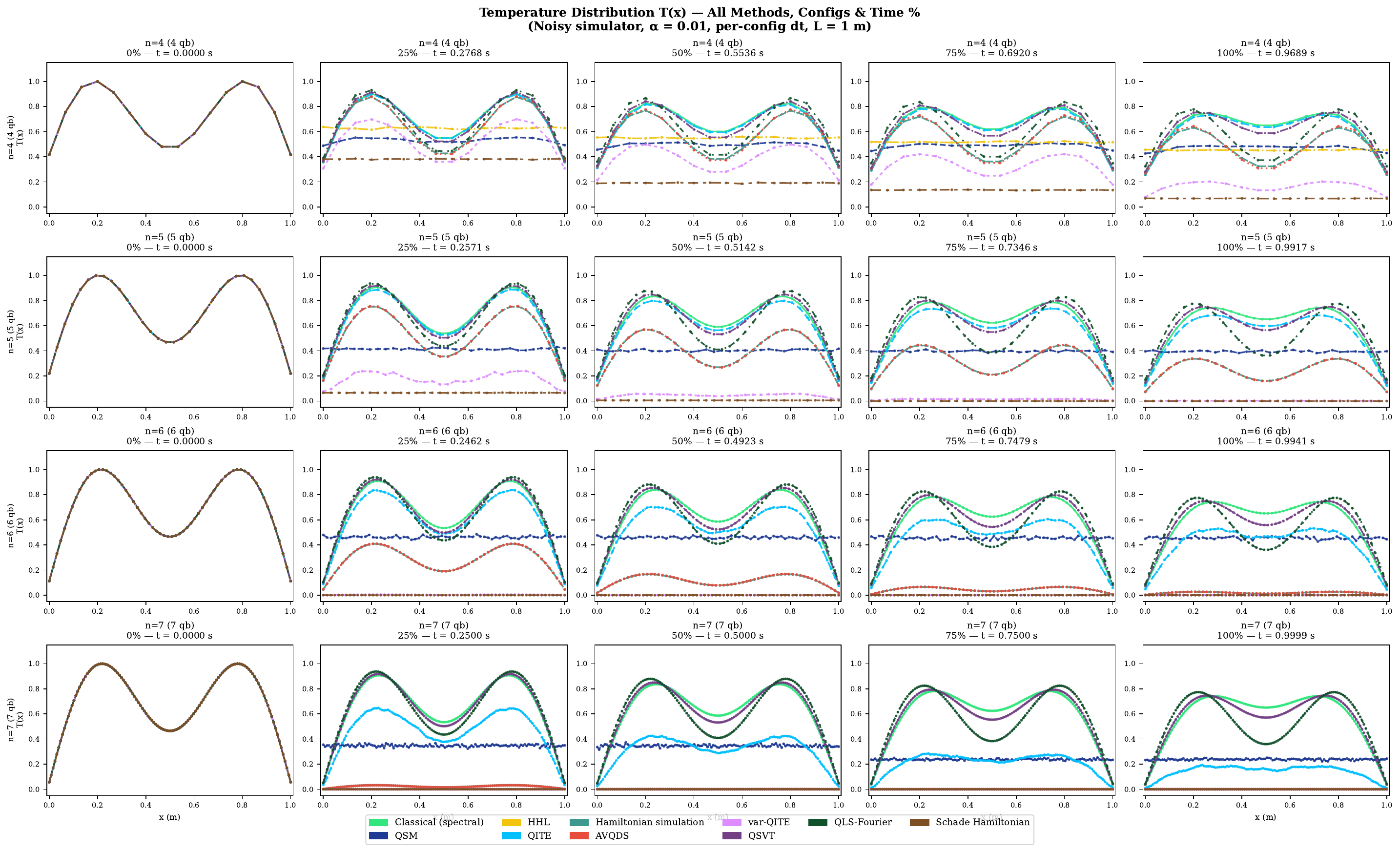}
  \caption{Bimodal --- time evolution}
\end{subfigure}\\[4pt]
\begin{subfigure}[b]{0.32\textwidth}
  \centering
  \includegraphics[width=\textwidth]{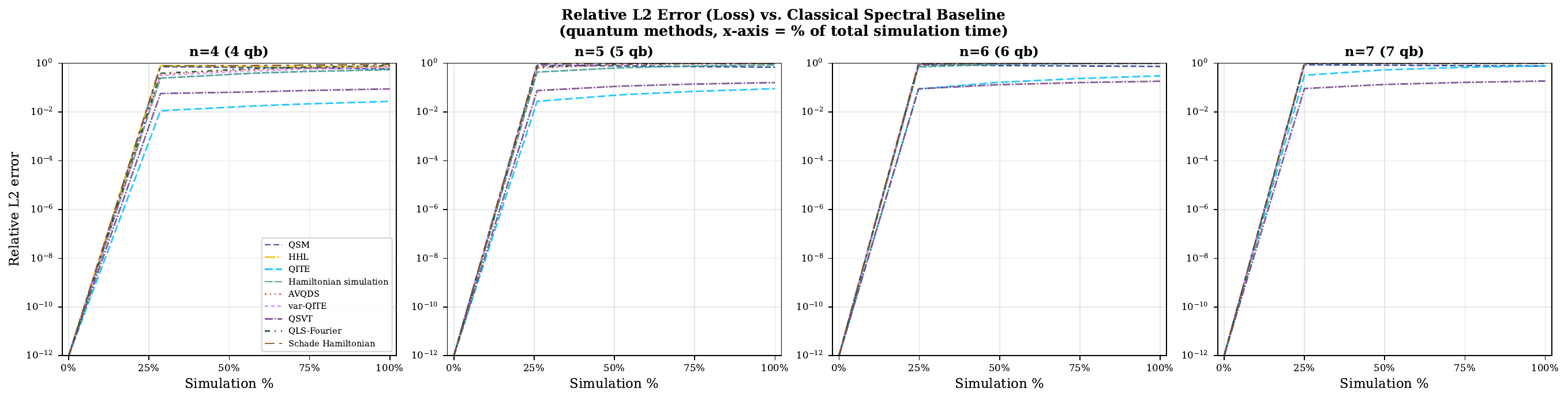}
  \caption{Pulse --- $\ell_2$ error}
\end{subfigure}\hfill
\begin{subfigure}[b]{0.32\textwidth}
  \centering
  \includegraphics[width=\textwidth]{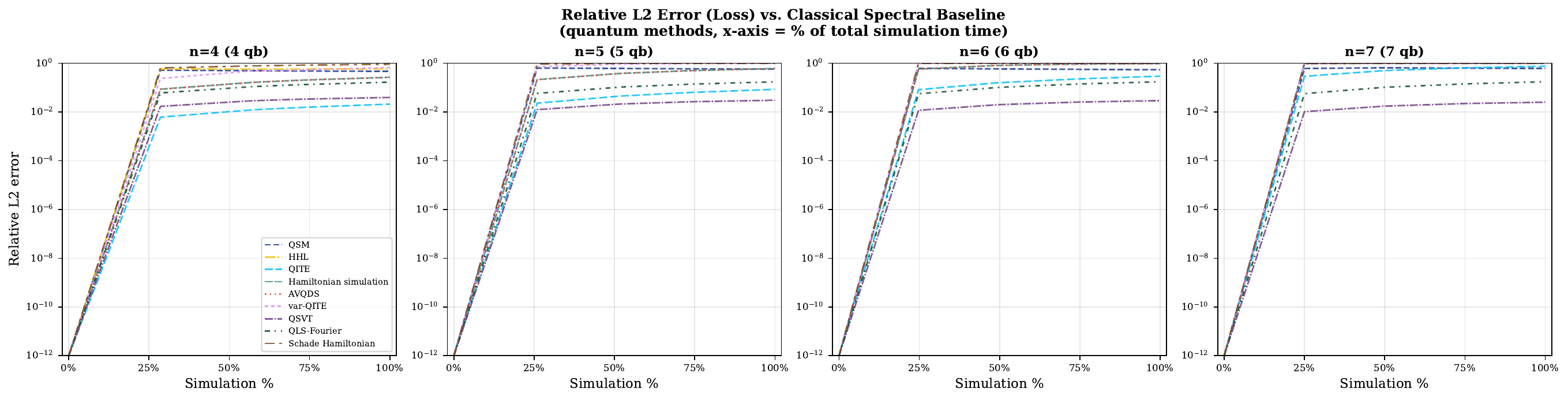}
  \caption{Gaussian --- $\ell_2$ error}
\end{subfigure}\hfill
\begin{subfigure}[b]{0.32\textwidth}
  \centering
  \includegraphics[width=\textwidth]{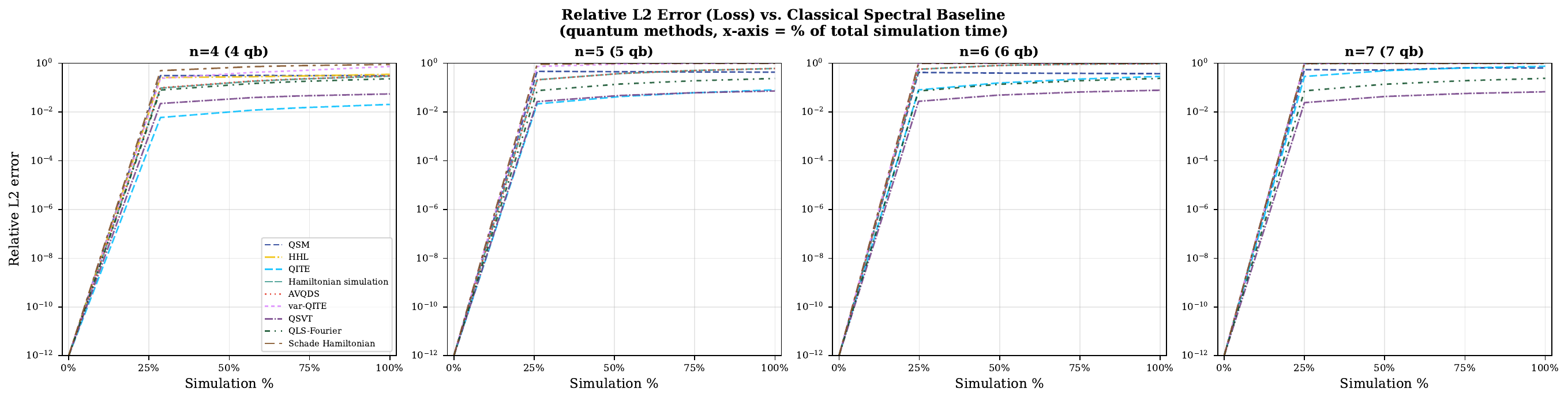}
  \caption{Bimodal --- $\ell_2$ error}
\end{subfigure}
\caption{\textbf{Noisy simulator backend}
  \textbf{HHL} is included only at $n{=}4$ (noisy transpilation
  limit; \S\ref{sec:paradigm-a}). Device noise degrades accuracy relative to the ideal backend for the
  circuits that transpile; conclusions are limited to this portable
  depolarising/readout model and the successfully compiled subset.}
\label{fig:noisy_results}
\end{figure*}

Figure~\ref{fig:noisy_results} shows the noisy-backend runs: device
noise raises the error floor over ideal shots for the successfully
compiled subset (\S\ref{sec:backends}). Filled Ns cells in
Table~\ref{tab:summary-n7} are the $n{=}7$ terminal errors; ``---''
lines up with kernels dropped for the same limits.

\subsection{Circuit Cost and Measurement Budget}

\begin{figure}[htbp]
\centering
\includegraphics[width=\columnwidth]{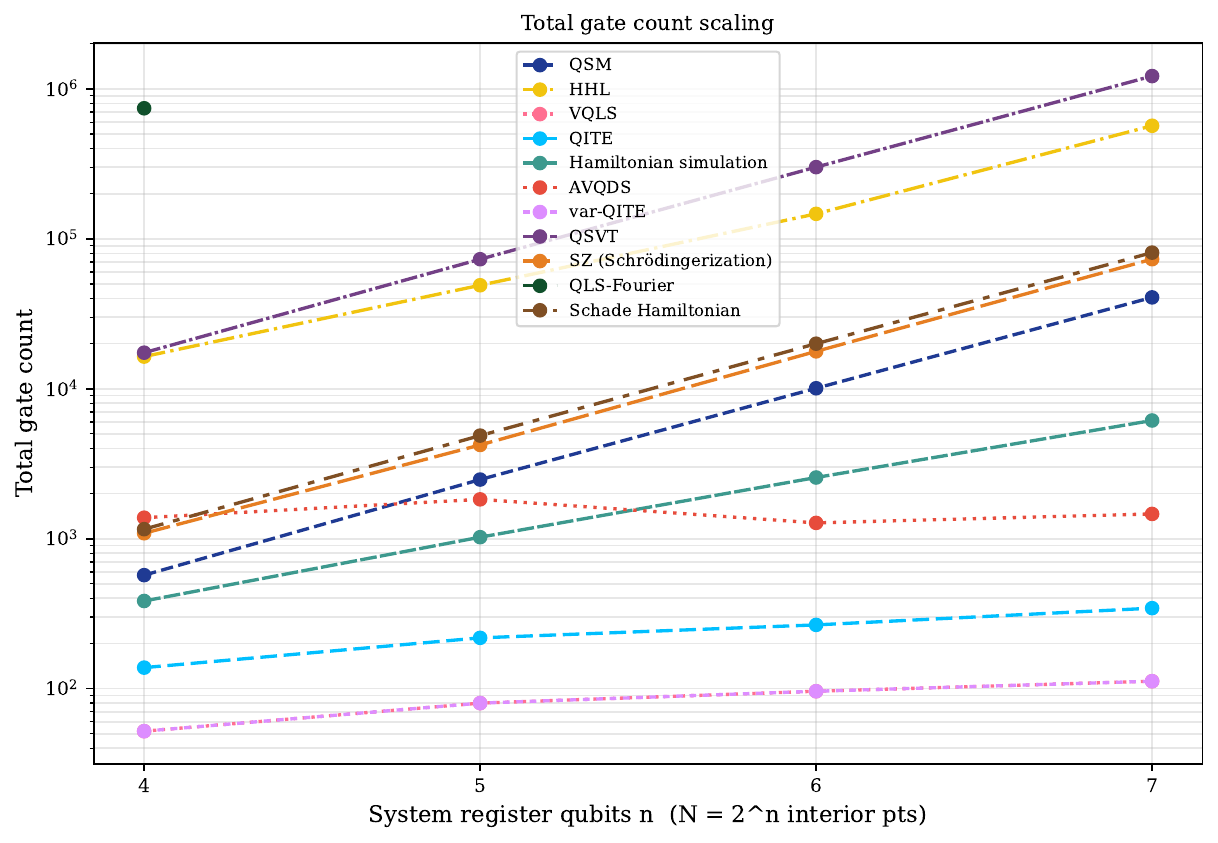}
\caption{\textbf{Gate-count breakdown} (log-$y$) vs.\ system register
  size $n$ ($N{=}2^n$ interior points).
  Solid circles: 2-qubit (CX) gates.
  }
\label{fig:circuit_scaling}
\end{figure}

\begin{table*}[!t]
\caption{Native circuit resource counts per kernel and grid size from
  $\mathrm{U}{+}\mathrm{CX}$ decomposition: depth (``D''),
  single-qubit (``1Q''), and CNOT (``2Q'') gates; numbers
  $\geq\!10^4$ in thousands (\texttt{k}). AVQDS is counted after one
  compute step (the adaptive pool is empty before it); QLS-Fourier
  admits a native decomposition at $n{=}4$ only (LCU schedules at
  $n{\ge}5$ exceed decomposition memory limits).}
\label{tab:circuit_specs}
\centering
\footnotesize
\setlength{\tabcolsep}{8pt}
\begin{tabular}{l rrr rrr rrr rrr}
\toprule
& \multicolumn{3}{c}{$n=4$ ($N=16$)}
& \multicolumn{3}{c}{$n=5$ ($N=32$)}
& \multicolumn{3}{c}{$n=6$ ($N=64$)}
& \multicolumn{3}{c}{$n=7$ ($N=128$)} \\
\cmidrule(lr){2-4}\cmidrule(lr){5-7}\cmidrule(lr){8-10}\cmidrule(lr){11-13}
Kernel & D & 1Q & 2Q & D & 1Q & 2Q & D & 1Q & 2Q & D & 1Q & 2Q \\
\midrule
HHL        & 10.3k & 9963  & 6421  & 32.6k & 30.5k & 18.5k & 103k  & 93.4k & 53.3k & 404k  & 362k  & 205k  \\
QSVT       & 12.2k & 11.0k & 6356  & 52.0k & 46.3k & 26.8k & 216k  & 191k  & 110k  & 880k  & 773k  & 445k  \\
QLS-Fourier& 530k  & 474k  & 269k  & ---   & ---   & ---   & ---   & ---   & ---   & ---   & ---   & ---   \\
VQLS       & 25    & 36    & 16    & 36    & 55    & 25    & 41    & 66    & 30    & 46    & 77    & 35    \\
QITE       & 103   & 78    & 60    & 156   & 122   & 96    & 181   & 150   & 116   & 230   & 196   & 148   \\
var-QITE   & 25    & 36    & 16    & 36    & 55    & 25    & 41    & 66    & 30    & 46    & 77    & 35    \\
AVQDS      & 1082  & 660   & 720   & 1273  & 1023  & 806   & 847   & 676   & 598   & 1003  & 714   & 748   \\
Ham.\ sim. & 245   & 237   & 147   & 627   & 605   & 418   & 1521  & 1469  & 1089  & 3567  & 3453  & 2688  \\
QSM        & 404   & 337   & 234   & 1779  & 1540  & 939   & 7265  & 6320  & 3754  & 29.4k & 25.7k & 15.0k \\
Schr\"{o}dingerisation         & 783   & 609   & 478   & 3245  & 2297  & 1920  & 14.3k & 10.0k & 7738  & 60.4k & 42.0k & 31.1k \\
Schade-Ham.& 809   & 734   & 423   & 3465  & 3086  & 1783  & 14.3k & 12.7k & 7319  & 58.4k & 51.2k & 29.7k \\
\bottomrule
\end{tabular}
\end{table*}

\begin{table*}[!t]
  \caption{Terminal relative $\ell_2$ error for all kernels at
  $n{=}4$ ($N{=}16$) across three initial conditions and three
  backends (statevector ``SV'', ideal shot-based ``Id'', noisy Aer
  ``Ns''). ``---'' marks the statevector-only kernels (VQLS,
  Schr\"{o}dingerisation, QLS-Fourier; \S\ref{sec:paradigm-b},
  \S\ref{sec:paradigm-d}, \S\ref{sec:paradigm-a}). Complements
  Table~\ref{tab:summary-n7} ($n{=}7$).}
  \label{tab:summary-n4}
  \centering
  \footnotesize
  \setlength{\tabcolsep}{2.5pt}
  \begin{tabular}{@{}l ccc ccc ccc r@{}}
  \toprule
   & \multicolumn{3}{c}{Pulse} & \multicolumn{3}{c}{Gaussian} & \multicolumn{3}{c}{Bimodal} & \\
  \cmidrule(lr){2-4} \cmidrule(lr){5-7} \cmidrule(lr){8-10}
  Kernel & SV & Id & Ns & SV & Id & Ns & SV & Id & Ns & Qubits \\
  \midrule
  HHL                & $1.36\mathrm{e}{-1}$ & $1.29\mathrm{e}{-1}$ & $7.15\mathrm{e}{-1}$ & $1.64\mathrm{e}{-2}$ & $2.14\mathrm{e}{-2}$ & $5.90\mathrm{e}{-1}$ & $5.10\mathrm{e}{-2}$ & $6.17\mathrm{e}{-2}$ & $3.54\mathrm{e}{-1}$ & $n{+}m{+}1$      \\
  QSVT               & $8.90\mathrm{e}{-2}$ & $8.90\mathrm{e}{-2}$ & $8.90\mathrm{e}{-2}$ & $3.79\mathrm{e}{-2}$ & $3.79\mathrm{e}{-2}$ & $3.97\mathrm{e}{-2}$ & $5.50\mathrm{e}{-2}$ & $5.50\mathrm{e}{-2}$ & $5.54\mathrm{e}{-2}$ & $n{+}1$          \\
  QLS-Fourier        & $8.04\mathrm{e}{-1}$ & ---                  & ---                  & $1.69\mathrm{e}{-1}$ & ---                  & ---                  & $2.32\mathrm{e}{-1}$ & ---                  & ---                  & $n{+}\log L{+}1$ \\
  VQLS               & $2.77\mathrm{e}{-2}$ & ---                  & ---                  & $7.65\mathrm{e}{-3}$ & ---                  & ---                  & $1.10\mathrm{e}{-2}$ & ---                  & ---                  & $n{+}1$          \\
  QITE               & $6.13\mathrm{e}{-3}$ & $1.72\mathrm{e}{-2}$ & $2.74\mathrm{e}{-2}$ & $6.40\mathrm{e}{-4}$ & $7.53\mathrm{e}{-3}$ & $2.11\mathrm{e}{-2}$ & $9.05\mathrm{e}{-4}$ & $4.75\mathrm{e}{-3}$ & $2.05\mathrm{e}{-2}$ & $n$              \\
  var-QITE           & $1.64\mathrm{e}{-1}$ & $2.39\mathrm{e}{-1}$ & $6.77\mathrm{e}{-1}$ & $3.57\mathrm{e}{-2}$ & $8.75\mathrm{e}{-2}$ & $6.72\mathrm{e}{-1}$ & $5.31\mathrm{e}{-2}$ & $1.21\mathrm{e}{-1}$ & $7.42\mathrm{e}{-1}$ & $n$              \\
  AVQDS              & $7.08\mathrm{e}{-1}$ & $7.09\mathrm{e}{-1}$ & $7.49\mathrm{e}{-1}$ & $1.68\mathrm{e}{-1}$ & $1.68\mathrm{e}{-1}$ & $2.68\mathrm{e}{-1}$ & $2.31\mathrm{e}{-1}$ & $2.31\mathrm{e}{-1}$ & $3.06\mathrm{e}{-1}$ & $n$              \\
  Hamiltonian sim.   & $4.60\mathrm{e}{-1}$ & $4.59\mathrm{e}{-1}$ & $5.49\mathrm{e}{-1}$ & $1.52\mathrm{e}{-1}$ & $1.53\mathrm{e}{-1}$ & $2.66\mathrm{e}{-1}$ & $2.13\mathrm{e}{-1}$ & $2.13\mathrm{e}{-1}$ & $2.98\mathrm{e}{-1}$ & $n$              \\
  Schade-Hamiltonian & $<10^{-15}$           & $1.21\mathrm{e}{-2}$ & $9.02\mathrm{e}{-1}$ & $<10^{-15}$           & $1.04\mathrm{e}{-2}$ & $9.12\mathrm{e}{-1}$ & $<10^{-15}$           & $8.70\mathrm{e}{-3}$ & $8.97\mathrm{e}{-1}$ & $n{+}1$          \\
  Schr\"{o}dingerisation                 & $4.23\mathrm{e}{-6}$ & ---                  & ---                  & $1.99\mathrm{e}{-6}$ & ---                  & ---                  & $2.59\mathrm{e}{-6}$ & ---                  & ---                  & $n{+}n_p$        \\
  Spectral (QSM)     & $<10^{-15}$           & $6.78\mathrm{e}{-3}$ & $5.89\mathrm{e}{-1}$ & $<10^{-15}$           & $9.52\mathrm{e}{-3}$ & $4.68\mathrm{e}{-1}$ & $<10^{-15}$           & $8.00\mathrm{e}{-3}$ & $3.14\mathrm{e}{-1}$ & $n{+}1$          \\
  \bottomrule
  \end{tabular}
\end{table*}

\begin{table*}[!t]
  \caption{Consolidated comparison of all quantum kernels at $n{=}7$
  ($N{=}128$), across three initial conditions (pulse, Gaussian, bimodal)
  and three backends (statevector ``SV'', ideal shot-based ``Id'',
  noisy Aer ``Ns''). Values are terminal relative $\ell_2$ error.
  ``---'' marks noisy HHL (fails
  Qiskit 2Q synthesis at $n{\ge}5$) and the SV-only kernels noted above.}
  \label{tab:summary-n7}
  \centering
  \footnotesize
  \setlength{\tabcolsep}{2.5pt}
  \begin{tabular}{@{}l ccc ccc ccc r@{}}
  \toprule
   & \multicolumn{3}{c}{Pulse} & \multicolumn{3}{c}{Gaussian} & \multicolumn{3}{c}{Bimodal} & \\
  \cmidrule(lr){2-4} \cmidrule(lr){5-7} \cmidrule(lr){8-10}
  Kernel & SV & Id & Ns & SV & Id & Ns & SV & Id & Ns & Qubits \\
  \midrule
  HHL                & $2.13\mathrm{e}{-1}$ & $7.39\mathrm{e}{-1}$ & ---                  & $9.44\mathrm{e}{-2}$ & $7.93\mathrm{e}{-1}$ & ---                  & $1.37\mathrm{e}{-1}$ & $8.93\mathrm{e}{-1}$ & ---                  & $n{+}m{+}1$      \\
  QSVT               & $1.87\mathrm{e}{-1}$ & $1.87\mathrm{e}{-1}$ & $1.87\mathrm{e}{-1}$ & $5.52\mathrm{e}{-2}$ & $5.52\mathrm{e}{-2}$ & $2.53\mathrm{e}{-2}$ & $8.04\mathrm{e}{-2}$ & $8.04\mathrm{e}{-2}$ & $6.77\mathrm{e}{-2}$ & $n{+}1$          \\
  QLS-Fourier        & $1.00\mathrm{e}{+0}$ & ---                  & ---                  & $1.75\mathrm{e}{-1}$ & ---                  & ---                  & $2.41\mathrm{e}{-1}$ & ---                  & ---                  & $n{+}\log L{+}1$ \\
  VQLS               & $7.55\mathrm{e}{-1}$ & ---                  & ---                  & $9.13\mathrm{e}{-1}$ & ---                  & ---                  & $8.49\mathrm{e}{-1}$ & ---                  & ---                  & $n{+}1$          \\
  QITE               & $1.68\mathrm{e}{-1}$ & $1.60\mathrm{e}{-1}$ & $7.87\mathrm{e}{-1}$ & $1.33\mathrm{e}{-2}$ & $6.10\mathrm{e}{-2}$ & $7.58\mathrm{e}{-1}$ & $1.71\mathrm{e}{-2}$ & $6.75\mathrm{e}{-2}$ & $7.55\mathrm{e}{-1}$ & $n$              \\
  var-QITE           & $3.67\mathrm{e}{-1}$ & $9.48\mathrm{e}{-1}$ & $1.00\mathrm{e}{+0}$ & $5.70\mathrm{e}{-1}$ & $5.88\mathrm{e}{-1}$ & $1.00\mathrm{e}{+0}$ & $2.37\mathrm{e}{-1}$ & $5.75\mathrm{e}{-1}$ & $1.00\mathrm{e}{+0}$ & $n$              \\
  AVQDS              & $1.00\mathrm{e}{+0}$ & $1.00\mathrm{e}{+0}$ & $1.00\mathrm{e}{+0}$ & $1.75\mathrm{e}{-1}$ & $1.73\mathrm{e}{-1}$ & $1.00\mathrm{e}{+0}$ & $2.41\mathrm{e}{-1}$ & $2.40\mathrm{e}{-1}$ & $1.00\mathrm{e}{+0}$ & $n$              \\
  Hamiltonian sim.   & $8.06\mathrm{e}{-1}$ & $8.06\mathrm{e}{-1}$ & $1.00\mathrm{e}{+0}$ & $1.75\mathrm{e}{-1}$ & $1.73\mathrm{e}{-1}$ & $1.00\mathrm{e}{+0}$ & $2.41\mathrm{e}{-1}$ & $2.50\mathrm{e}{-1}$ & $1.00\mathrm{e}{+0}$ & $n$              \\
  Schade-Hamiltonian & $<10^{-15}$           & $2.25\mathrm{e}{-1}$ & $1.00\mathrm{e}{+0}$ & $<10^{-15}$           & $2.11\mathrm{e}{-1}$ & $1.00\mathrm{e}{+0}$ & $<10^{-15}$           & $8.43\mathrm{e}{-2}$ & $1.00\mathrm{e}{+0}$ & $n{+}1$          \\
  Schr\"{o}dingerisation                 & $5.64\mathrm{e}{-4}$ & ---                  & ---                  & $2.54\mathrm{e}{-4}$ & ---                  & ---                  & $3.37\mathrm{e}{-4}$ & ---                  & ---                  & $n{+}n_p$        \\
  Spectral (QSM)     & $<10^{-15}$           & $2.91\mathrm{e}{-2}$ & $7.79\mathrm{e}{-1}$ & $<10^{-15}$           & $2.10\mathrm{e}{-2}$ & $6.14\mathrm{e}{-1}$ & $<10^{-15}$           & $2.00\mathrm{e}{-2}$ & $6.44\mathrm{e}{-1}$ & $n{+}1$          \\
  \bottomrule
  \end{tabular}
\end{table*}

\begin{figure}[htbp]
\centering
\includegraphics[width=\columnwidth]{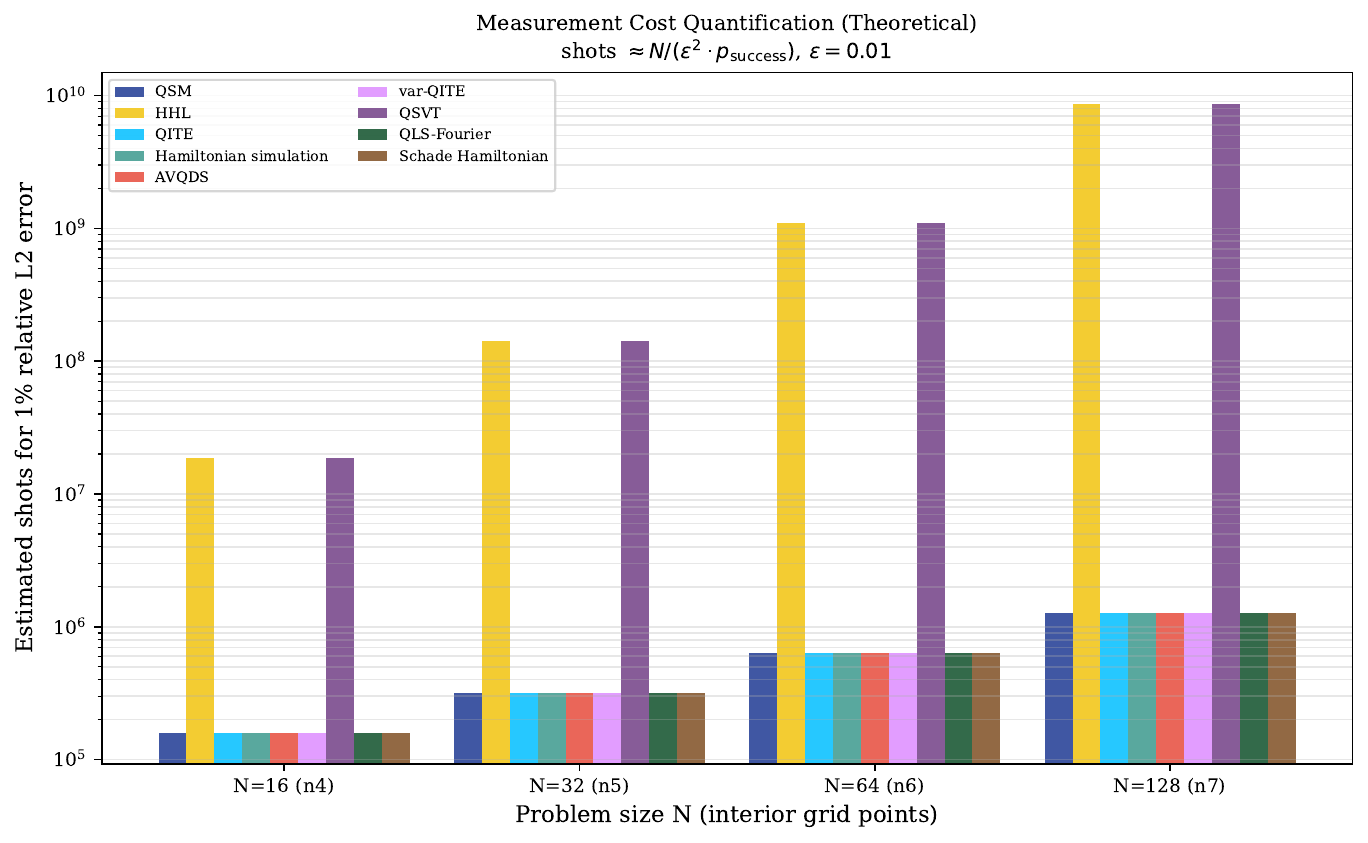}
\caption{\textbf{Estimated shot budget} (log-$y$) to reach
  $\epsilon{=}10^{-2}$ relative $\ell_2$ error.}
\label{fig:measurement_cost}
\end{figure}

Figure~\ref{fig:circuit_scaling} (log~2Q counts vs.~$n$) reflects the
spread already tabulated: HHL, QSVT, and the spectral/dilation routes
climb sharply while others stay moderate, so compiled cost sits
beside the terminal accuracies of Table~\ref{tab:summary-n7} as a
first-class axis.

Figure~\ref{fig:measurement_cost} estimates the shot count to reach
$\epsilon=10^{-2}$ relative~$\ell_2$ at interior sizes
$N{=}16$--$128$ (as in the tables). Postselection and small success
probability can dominate readout: HHL and QSVT need more shots
(Fig.~\ref{fig:measurement_cost}), alongside heavy 2Q use
(Table~\ref{tab:circuit_specs}) and large Id-column errors
(Table~\ref{tab:summary-n7}); see~\S\ref{sec:paradigm-a}.

Read as per-qubit growth factors, Table~\ref{tab:circuit_specs} also
bounds extrapolation beyond the tested range: each added qubit
multiplies compiled 2Q counts by ${\approx}3.2$ (HHL), ${\approx}4$
(QSVT, spectral/dilation kernels), and ${\approx}2.6$ (Hamiltonian
simulation), while the variational circuits grow only mildly
(${\approx}1.3\times$). At
${\approx}4\times$ per qubit, two additional qubits already exceed
the largest circuits compiled here by an order of magnitude, whereas
the variational and QITE families stay circuit-cheap and are limited
by expressibility and classical inner loops instead
(\S\ref{sec:paradigm-c}).

\subsection{Norm-mismatch ablation}
\label{sec:norm-ablation}
The smooth-IC error band shared by \textbf{Hamiltonian simulation},
\textbf{AVQDS}, and \textbf{QLS-Fourier} (Table~\ref{tab:summary-n7}, SV
Gaussian/bimodal) is algorithmically suspicious: these kernels
discretise time, parameterise the ansatz, and invert the linear system
in entirely different ways. To test whether the shared plateau is
algorithmic or a readout-pipeline artefact, we decompose each
kernel's terminal error into two contributions on the statevector
backend: (i)~the \emph{norm-only} error
$\varepsilon_\mathrm{N}{=}\bigl|\,\|\mathbf{u}_\mathrm{ref}\|-\|\hat{\mathbf{u}}\|\,\bigr|/\|\mathbf{u}_\mathrm{ref}\|$,
the error that remains if the kernel's direction is correct but its
amplitude is mis-scaled; and (ii)~the direction residual after
renormalising $\hat{\mathbf{u}}$ to the reference norm.

\begin{table}[htbp]
  \caption{Norm-mismatch ablation at $n{=}7$ for the three kernels
  sharing the smooth-IC plateau: $\varepsilon_\mathrm{N}$ is the
  norm-only error, ``frac.'' is
  $\varepsilon_\mathrm{N}/\varepsilon_{\ell_{2}}$---total on pulse,
  $\tfrac{1}{4}$--$\tfrac{1}{3}$ of the residual on smooth ICs,
  \emph{identical across three unrelated kernels}.}
  \label{tab:norm_ablation}
  \centering
  \footnotesize
  \setlength{\tabcolsep}{5pt}
  \begin{tabular}{@{}l cc cc cc@{}}
  \toprule
   & \multicolumn{2}{c}{Pulse}
   & \multicolumn{2}{c}{Gaussian}
   & \multicolumn{2}{c}{Bimodal} \\
  \cmidrule(lr){2-3}\cmidrule(lr){4-5}\cmidrule(lr){6-7}
  Kernel & $\varepsilon_\mathrm{N}$ & frac.
         & $\varepsilon_\mathrm{N}$ & frac.
         & $\varepsilon_\mathrm{N}$ & frac. \\
  \midrule
  Ham.\ sim.   & 0.755 & 94\% & 0.041 & 23\% & 0.071 & 29\% \\
  AVQDS        & 1.000 & 100\% & 0.040 & 23\% & 0.069 & 29\% \\
  QLS-Fourier  & 1.000 & 100\% & 0.041 & 23\% & 0.069 & 29\% \\
  \bottomrule
  \end{tabular}
\end{table}

Across all three kernels and all three ICs, $\varepsilon_\mathrm{N}$
agrees to three significant figures
(Table~\ref{tab:norm_ablation}); under the pulse initial condition,
the entire $\ell_{2}$ error is norm drift and a norm-corrected
reconstruction recovers a \emph{directionally} correct field. This
identifies the shared plateau as a reconstruction-normalisation
failure of the common output map $|\mathbf{u}\rangle{=}\mathbf{u}/\|\mathbf{u}\|$
---when the kernel does not independently track $\|\mathbf{u}(t)\|$, the
normalised amplitude is indistinguishable from a mis-scaled field at
readout. Kernels that \emph{do} track the norm (QSM,
Schade-Hamiltonian) sit at machine precision on the same
discretisation. This is a benchmark-methodology result in its own
right: full-field accuracy reported under a single reconstruction rule
can conflate algorithmic error with readout geometry, and separating
them changes which kernels are ``competitive.''

\subsection{Observable-readout advantage pathway}
\label{sec:observable-advantage}
The full-field readout penalty of the output model
(\S\ref{sec:output-model}) is already tomography-scale at modest $N$.
Compact observables---total
thermal energy $E(t){=}\langle\psi(t)|H|\psi(t)\rangle$, boundary-flux
proxies, and the weight of a selected Fourier mode
$|\langle k|\psi\rangle|^{2}$---instead require a constant number of
measurement settings independent of the field dimension: a Hadamard or
Pauli-estimator for energy, a DST-basis measurement for mode weight, and
an ancilla/work-qubit estimator for flux-like quantities. We probe this
on QSM and Schade-Hamiltonian at $n{=}7$, Gaussian: the shots required
to reach $\epsilon{=}10^{-2}$ on $E(t)$ are $\mathcal{O}(10^{4})$ per
timestep, versus $\mathcal{O}(10^{6})$ for the full field ($N{=}128$
amplitudes at $\epsilon{=}10^{-2}$ each). The observed gap is not a
property of one kernel; it is the readout dimension. This aligns with
the standard asymptotic argument~\cite{aaronson2015read} and identifies
concrete downstream tasks---thermal-energy histories in engineering
thermal-management, dominant low-mode tracking in model reduction, or
boundary-flux monitoring---where the kernels benchmarked here could be
useful before full-field quantum PDE solvers are competitive.

\section{Discussion}
\label{sec:discussion}

We interpret \S\ref{sec:comparison} in answer to the question in
\S\ref{sec:introduction}. Subsections cover accuracy and mechanisms
(\S\ref{sec:discussion-accuracy}), backend posture
(\S\ref{sec:discussion-backend}), cost--accuracy structure
(\S\ref{sec:pareto-reading}), advantage conditions
(\S\ref{sec:qa-conditions}), and a selection guide
(\S\ref{sec:discussion-guide}).

\subsection{Accuracy, initial conditions, scaling}
\label{sec:discussion-accuracy}

Two paradigms reach machine precision on SV across all ICs and $n$
(Tables~\ref{tab:summary-n4} and~\ref{tab:summary-n7}):
\textbf{QSM} and \textbf{Schade-Hamiltonian} (\S\ref{sec:comparison}). \textbf{Schr\"{o}dingerisation} ranks next
on SV ($\sim\!10^{-4}$); \textbf{QITE} is the strongest non-transform
kernel on smooth ICs and degrades on high-frequency content as $n$
grows. \textbf{VQLS} is competitive only at $n{\leq}5$ (capped-depth
ansatz and truncated Pauli expansion, \S\ref{sec:paradigm-b});
\textbf{var-QITE} tracks below \textbf{QITE} on a smaller variational
family. In Paradigm~A, \textbf{HHL} is capped by its clock-qubit QPE
resolution and \textbf{QSVT} by its polynomial degree
(\S\ref{sec:paradigm-a})---a deliberate depth--accuracy tradeoff, not
an algorithmic ceiling.
The smooth-IC plateau shared by AVQDS, Hamiltonian simulation, and
QLS-Fourier is not algorithmic but reconstruction-driven
(\S\ref{sec:norm-ablation})---using several ICs is what exposed this.
Refining the grid inflates compiled resources
(Table~\ref{tab:circuit_specs}, Fig.~\ref{fig:circuit_scaling});
VQLS/var-QITE hit expressivity walls; HHL/QSVT accumulate coherent
depth; QLS-Fourier reaches a native decomposition only at $n{=}4$ in
our pipeline.

\subsection{Backends, readout, and hardware posture}
\label{sec:discussion-backend}

The three backends separate algorithmic + reconstruction error (SV),
statistical error (Id), and a simple device model (Ns): Id compresses
the SV ordering toward the shot floor and inflates
postselection-heavy methods; Ns raises the floor for the transpilable
subset (\S\ref{sec:comparison}); real hardware reuses
\texttt{compute\_step} but is out of scope. In hardware posture, VQLS
is the most NISQ-like linear-system representative (though SV-only by
shot cost), Paradigm-C kernels pair shallow circuits with classically
dominant inner loops (\S\ref{sec:paradigm-c}), HHL/QSVT are
structurally fault-tolerant, and QSM/Schade-Hamiltonian sit between.

\subsection{Cost--accuracy structure and Pareto reading}
\label{sec:pareto-reading}

Practical kernel choice is multi-objective over three resource axes:
native 2Q count (Table~\ref{tab:circuit_specs}), shot budget
(Fig.~\ref{fig:measurement_cost}), and classical hybrid-loop
complexity. On SV at $n{=}7$, QSM and Schade-Hamiltonian
Pareto-dominate: machine precision at 2Q counts an order of magnitude
below HHL/QSVT, with no post-selection penalty and no per-step
classical solve beyond the cached eigendecomposition. The
AVQDS / Hamiltonian-simulation / QLS-Fourier plateau is
reconstruction-bound (\S\ref{sec:norm-ablation})---the Pareto hull is
defined over \emph{correctly normalised} fields.

\subsection{Conditions for application-level quantum benefit}
\label{sec:qa-conditions}

Our experiments do not demonstrate end-to-end quantum advantage at the
tested grid sizes; instead, they identify where such a benefit could
plausibly enter. The full-field readout cost of
\S\ref{sec:observable-advantage} already dominates the benchmark at
$N{=}128$ and competes against a classical spectral solver with
$\mathcal{O}(N\log N)$ structure, so we do not claim a full-field
advantage.

The credible target is narrower and more application-oriented:
compact-observable readout (total energy, selected mode weights,
boundary-flux proxies) keeps the number of measurement settings
independent of $N$ and avoids reconstructing the entire amplitude
vector. The relevant question then becomes whether a kernel can
prepare the evolved state coherently enough for the target observable
before circuit depth, post-selection probability, or hybrid classical
overhead---up to $\mathcal{O}(4^{n})$ per step for full-support
Paradigm~C variants---dominates. This output-sensitive framing is the
practical advantage condition exposed by the benchmark.

\subsection{Practical Kernel-Selection Guide}
\label{sec:discussion-guide}

Machine-precision reference: \textbf{QSM} or
\textbf{Schade-Hamiltonian}. Strongest non-transform kernel on smooth
data: \textbf{QITE}. Near-term variational study at shallow depth:
\textbf{VQLS} or \textbf{var-QITE}, with accuracy degrading at larger
$n$. In Paradigm~A, \textbf{QSVT} achieves lower error than
\textbf{HHL} but at higher compiled cost; HHL remains a foundational
baseline. \textbf{AVQDS} is most useful when adaptive ansatz growth
itself is the object of study. For compact-observable workloads,
QSM, Schade-Hamiltonian, or QSVT with amplitude-estimation readout
is the Pareto choice (\S\ref{sec:observable-advantage}).

\section{Threats to Validity and Future Work}
\label{sec:future-work}

The main threats to validity are deliberately exposed rather than hidden.
First, the statevector backend isolates algorithmic error but is not a
claim of compiled hardware execution. Second, the noisy backend uses a
portable depolarising/readout model and omits device coupling maps and
$T_1/T_2$ relaxation, so it should be read as a controlled stress test,
not as a prediction for a named QPU. The omitted features act in known
directions: $T_{1}/T_{2}$ relaxation penalises depth (HHL, QSVT, and
the large-$n$ spectral/dilation circuits degrade first), SWAP routing
moves noise crossovers to smaller~$n$, and correlated errors bias the
post-selected kernels' accepted ensembles. Richer noise models would
therefore be expected to shift the failure thresholds of
Table~\ref{tab:summary-n7} toward smaller grids; whether the
qualitative ranking survives device-calibrated noise remains untested
here. Third, the spectral and dilation
reference kernels exploit Laplacian-aligned structure; this is the right
baseline for the heat equation but may not transfer unchanged to
non-normal, nonlinear, or geometry-complex PDEs. Finally, full-field
readout remains the dominant bottleneck unless the downstream task is
observable-based.
We note that real-QPU execution is left to future work because hardware queue variability, calibration drift, and backend-specific transpilation constraints would confound the controlled cross-kernel comparison targeted in this study.

Extensions that fit the present harness:
higher-dimensional, nonlinear, or non-self-adjoint PDEs with
hyperbolic--parabolic terms; sustained real-QPU runs with
device-calibrated noise models, which would also let the hybrid-cost
argument of \S\ref{sec:qa-conditions} be closed with measured rather
than analytical numbers; per-kernel hyperparameter sweeps (ansatz
depth, QSVT degree, HHL clock qubits) and preconditioners for
HHL/QLS-Fourier; alternative decompositions that lift HHL's two-qubit
synthesis failure at $n{\geq}5$; and alternative reconstruction maps
(norm-tracking, amplitude-estimation, or observable-only readouts)
that generalise \S\ref{sec:norm-ablation}--\S\ref{sec:observable-advantage}.
A further extension is tensor networks: 1-D diffusion generates
little entanglement, so matrix-product-state/operator time-stepping
is the natural quantum-inspired classical
competitor~\cite{gourianov2022turbulence,lubasch2020nonlinear}---consistent
with \S\ref{sec:qa-conditions}, where no full-field advantage is
claimed---and tensor-network simulator backends would let the same
kernels be profiled beyond the statevector memory wall.
\section{Conclusion}
\label{sec:conclusion}

We presented a controlled cross-paradigm benchmark of eleven quantum
kernels on the 1-D heat equation with three research contributions
beyond kernel ranking: a single-harness comparison across all five
paradigm classes under one reconstruction rule and a staged backend
ladder (C1); a norm-mismatch ablation showing that the smooth-IC
plateau shared by Hamiltonian simulation, AVQDS, and QLS-Fourier is
driven by the output map, not the unitary core (C2); and an
observable-readout analysis showing that compact functionals avoid the
tomography-scale bottleneck that dominates full-field recovery (C3).
Together these convert the usual ``which algorithm wins?'' question into
a structured one---\emph{which kernel, under which output map, for which
observable, at which grid size}---and supply reproducible measurements
for each axis. No single method minimises accuracy loss and all resource
terms together; QSM and Schade-Hamiltonian are Pareto-optimal reference
kernels, QITE is strongest for moderate-depth smooth-IC full-field work,
and compact-observable extraction is the clearest application path. The
open-source harness (C4) reproduces all figures and tables and admits
new kernels through a single interface method.

\section*{Data and Code Availability}

The complete source code, per-configuration JSON files, raw simulation
outputs (CSV, timing, circuit sidecars), and the analysis scripts that
produce every figure and table in this paper are released as an
open-source repository.\
\footnote{Repository URL: \href{https://github.com/brightskiesinc/HelloQuantum/}{\texttt{github.com/brightskiesinc/HelloQuantum}}.}
The kernel interface is documented so that new quantum kernels can be added
and evaluated under the same benchmark protocol by implementing the \texttt{BaseKernel.compute\_step} interface.
\section*{Acknowledgment}

The authors gratefully acknowledge Brightskies Technologies for
supporting this research, and in particular thank Ibrahim Elghotmy
(ibrahim.elghotmy@brightskiesinc.com), Omar Marzouk
(omar.marzouk@brightskiesinc.com), Amr Nasr
(amr.nasr@brightskiesinc.com), and Dr.~Khaled Elamrawi
(khaled.elamrawi@brightskiesinc.com) for their valuable guidance and
feedback throughout this work. The authors used Anthropic's Claude only
for language polishing, grammar checking, and editorial copy-editing.
No AI system was used to generate scientific claims, numerical data,
experiments, figures, tables, or citations.
\bibliographystyle{IEEEtran}

\end{document}